\documentclass[preprint,12pt,3p]{elsarticle}

\usepackage{graphics}
\usepackage{graphicx}
\usepackage{epsfig}

\usepackage{amssymb}
\usepackage{xcolor}
\usepackage{times}
\usepackage{epsfig}
\usepackage{graphicx}
\usepackage{subfig}
\usepackage{float}
\usepackage{amsmath}
\usepackage{amssymb}
\usepackage{algorithm}
\usepackage{algorithmic}
\usepackage{multirow}
\usepackage{setspace}

\makeatletter \@fpsep = 2pt \makeatother
\begin{document}

\begin{spacing}{1.0}

\begin{frontmatter}

\title{Deep Learning on Image Denoising: An Overview}

\author[label1,label2]{Chunwei Tian}
\author[label3]{Lunke Fei}
\author[label4]{Wenxian Zheng}
\author[label1,label2,label5]{Yong Xu\corref{cor1}}
\ead{yongxu@ymail.com}
\author[label6,label5]{Wangmeng Zuo}
\author[label7]{Chia-Wen Lin}
\address[label1]{Bio-Computing Research Center,  Harbin Institute of Technology, Shenzhen, Shenzhen, 518055, Guangdong, China}
\address[label2]{Shenzhen Key Laboratory of Visual Object Detection and Recognition, Shenzhen, 518055, Guangdong, China}
\address[label3]{School of Computers, Guangdong University of Technology, Guangzhou, 510006, Guangdong, China}
\address[label4]{Tsinghua Shenzhen International Graduate School, Shenzhen, 518055, Guangdong, China}
\address[label5]{Peng Cheng Laboratory, Shenzhen, 518055, Guangdong, China}
\address[label6]{School of Computer Science and Technology, Harbin Institute of Technology, Harbin, 150001, Heilongjiang, China}
\address[label7]{Department of Electrical Engineering and the Institute of Communications Engineering, National Tsing Hua University, Hsinchu, Taiwan}
\cortext[cor1]{Corresponding author}

\begin{abstract}
Deep learning techniques have received much attention in the area of image denoising. However, there are substantial differences in the various types of deep learning methods dealing with image denoising. Specifically, discriminative learning based on deep learning can ably address the issue of Gaussian noise. Optimization models based on deep learning are effective in estimating the real noise. However, there has thus far been little related research to summarize the different deep learning techniques for image denoising. In this paper, we offer a comparative study of deep techniques in image denoising. We first classify the deep convolutional neural networks (CNNs) for additive white noisy images; the deep CNNs for real noisy images; the deep CNNs for blind denoising and the deep CNNs for hybrid noisy images, which represents the combination of noisy, blurred and low-resolution images. Then, we analyze the motivations and principles of the different types of deep learning methods. Next, we compare the state-of-the-art methods on public denoising datasets in terms of quantitative and qualitative analysis. Finally, we point out some potential challenges and directions of future research.
\end{abstract}

\begin{keyword}
Deep learning \sep Image denoising \sep Real noisy images \sep Blind denoising \sep Hybrid noisy images
\end{keyword}

\end{frontmatter}

\section{Introduction}
\label{sec-1}
Digital image devices have been widely applied in many fields, including recognition of individuals \cite{lei2016skin,wen2018incomplete,wen2020generalized}, and remote sensing \cite{du2019improved}. The captured image is a degraded image from the latent observation, in which the degradation processing is affected by factors such as lighting and noise corruption \cite{zhang2017image,zha2018rank}. Specifically, the noise is generated in the processes of transmission and compression from the unknown latent observation \cite{xu2018external}. It is essential to use image denoising techniques to remove the noise and recover the latent observation from the given degraded image.

Image denoising techniques have attracted much attention in recent 50 years \cite{bernstein1987adaptive,xu2015patch}. At the outset, nonlinear and non-adaptive filters were used for image applications \cite{huang1971stability}. Nonlinear filters can preserve the edge information to suppress  the noise, unlike linear filters \cite{pitas1986nonlinear}. Adaptive nonlinear filters depend on local signal-to-noise ratios to derive an appropriate weighting factor for removing noise from an image corrupted by the combination of additive random, signal-dependent, impulse noise and additive random noise \cite{bernstein1987adaptive}. Non-adaptive filters can simultaneously use edge information and signal-to-noise ratio information to estimate the noise \cite{hong2000edge}. In time, machine learning methods, such as sparse-based methods were successfully applied in image denoising \cite{dabov2007image}. A non-locally centralized sparse representation (NCSR) method used nonlocal self-similarity to optimize the sparse method, and obtained high performance for image denoising \cite{dong2012nonlocally}. To reduce computational costs, a dictionary learning method was used to quickly filter the noise \cite{elad2006image}. To recover the detailed information of the latent clean image, priori knowledge (i.e., total variation regularization) can smooth the noisy image in order to deal with the corrupted image \cite{osher2005iterative,ren2019simultaneous}. More competitive methods for image denoising can be found in \cite{mairal2009non,zuo2014gradient,zhang2017beyond}, including the Markov random field (MRF) \cite{schmidt2014shrinkage}, the weighted nuclear norm minimization (WNNM) \cite{gu2014weighted}, learned simultaneous sparse coding (LSSC) \cite{mairal2009non}, cascade of shrinkage fields (CSF) \cite{schmidt2014shrinkage}, trainable nonlinear reaction diffusion (TNRD) \cite{chen2016trainable} and gradient histogram estimation and preservation (GHEP) \cite{zuo2014gradient}.

Although most of the above methods have achieved reasonably good performance in image denoising, they suffered from several drawbacks \cite{lucas2018using}, including the need for optimization methods for the test phase, manual setting parameters, and a certain model for single denoising tasks. Recently, as architectures became more flexible, deep learning techniques gained the ability to overcome these drawbacks \cite{lucas2018using}.

The original deep learning technologies were first used in image processing in the 1980s \cite{fukushima1982neocognitron} and were first used in image denoising by Zhou et al. \cite{chiang1989multi,zhou1987novel}. That is, the proposed denoising work first used a neural network with both the known shift-invariant blur function and additive noise to recover the latent clean image. After that, the neural network used weighting factors to remove complex noise \cite{chiang1989multi}. To reduce the high computational costs, a feedforward network was proposed to make a tradeoff between denoising efficiency and performance \cite{tamura1989analysis}. The feedforward network can smooth the given corrupted image by Kuwahara filters, which were similar to convolutions. In addition, this research proved that the mean squared error (MSE) acted as a loss function and was not unique to neural networks \cite{de1999applicability,greenhill1994relative}. Subsequently, more optimization algorithms were used to accelerate the convergence of the trained network and to promote the denoising performance \cite{bedini1992image,de1992image,gardner1989training}. The combination of maximum entropy and prima-dual Lagrangian multipliers to enhance the expressive ability of neural networks proved to be a good tool for image denoising \cite{bedini1990neural}. To further make a tradeoff between fast execution and denoising performance, greedy algorithms and asynchronous algorithms were applied in neural networks \cite{paik1992image}. Alternatively, designing a novel network architecture proved to be very competitive in eliminating the noise, through either increasing the depth or changing activation function \cite{sivakumar1993image}. Cellular neural networks (CENNs) mainly used nodes with templates to obtain the averaging function and effectively suppress the noise \cite{sivakumar1993image,nossek1993special}. Although this proposed method can obtain good denoising results, it requires the parameters of the templates to be set manually. To resolve this problem, the gradient descent was developed \cite{zamparelli1997genetically,lee1996color}. To a certain degree, these deep techniques can improve denoising performance. However, these networks did not easily allow the addition of new plug-in units, which limited their applications in the real world \cite{fukushima1980neocognitron}.

Based on the reasons above, convolutional neural networks (CNNs) were proposed \cite{lo1995artificial,ren2020single}. The CNN as well as the LeNet had real-world application in handwritten digit recognition \cite{lecun1998gradient}. However, due to the following drawbacks, they were not widely applied in computer systems \cite{krizhevsky2012imagenet}. First, deep CNNs can generate vanishing gradients. Second, activation functions such as sigmoid \cite{marreiros2008population} and tanh \cite{jarrett2009best} resulted in high computational cost. Third, the hardware platform did not support the complex network. However, that changed in 2012 with AlexNet in that year's ImageNet Large-Scale Visual Recognition Challenge (ILSVRC) \cite{krizhevsky2012imagenet}. After that, deep network architectures (e.g., VGG \cite{simonyan2014very} and GoogLeNet \cite{szegedy2015going}) were widely applied in the fields of image \cite{wu2019dsn,wang2018multi,li2020similarity}, video \cite{liu2017deep,yuan2020visual}, nature language processing \cite{duan2018attention} and speech processing \cite{zhang2018deep}, especially low-level computer vision \cite{peng2019dilated,tian2019enhanced}.

Deep networks were first applied in image denoising in 2015 \cite{liang2015stacked,xu2015denoising}. The proposed network need not manually set parameters for removing the noise. After then, deep networks were widely applied in speech \cite{zhang2015deep}, video \cite{yuan2020learning} and image restoration \cite{tian2020coarse,ren2020brn}. Mao et al. \cite{mao2016image} used multiple convolutions and deconvolutions to suppress the noise and recover the high-resolution image. For addressing multiple low-level tasks via a model, a denoising CNN (DnCNN) \cite{zhang2017beyond} consisting of convolutions, batch normalization (BN) \cite{ioffe2015batch}, rectified linear unit (ReLU) \cite{nair2010rectified} and residual learning (RL) \cite{he2016deep} was proposed to deal with image denoising, super-resolution, and JPEG image deblocking. Taking into account the tradeoff between denoising performance and speed, a color non-local network (CNLNet) \cite{lefkimmiatis2017non} combined non-local self-similarity (NLSS) and CNN to efficiently remove color-image noise.
\begin{figure}[!htb]
\centering
\subfloat{\includegraphics[width=6.5in]{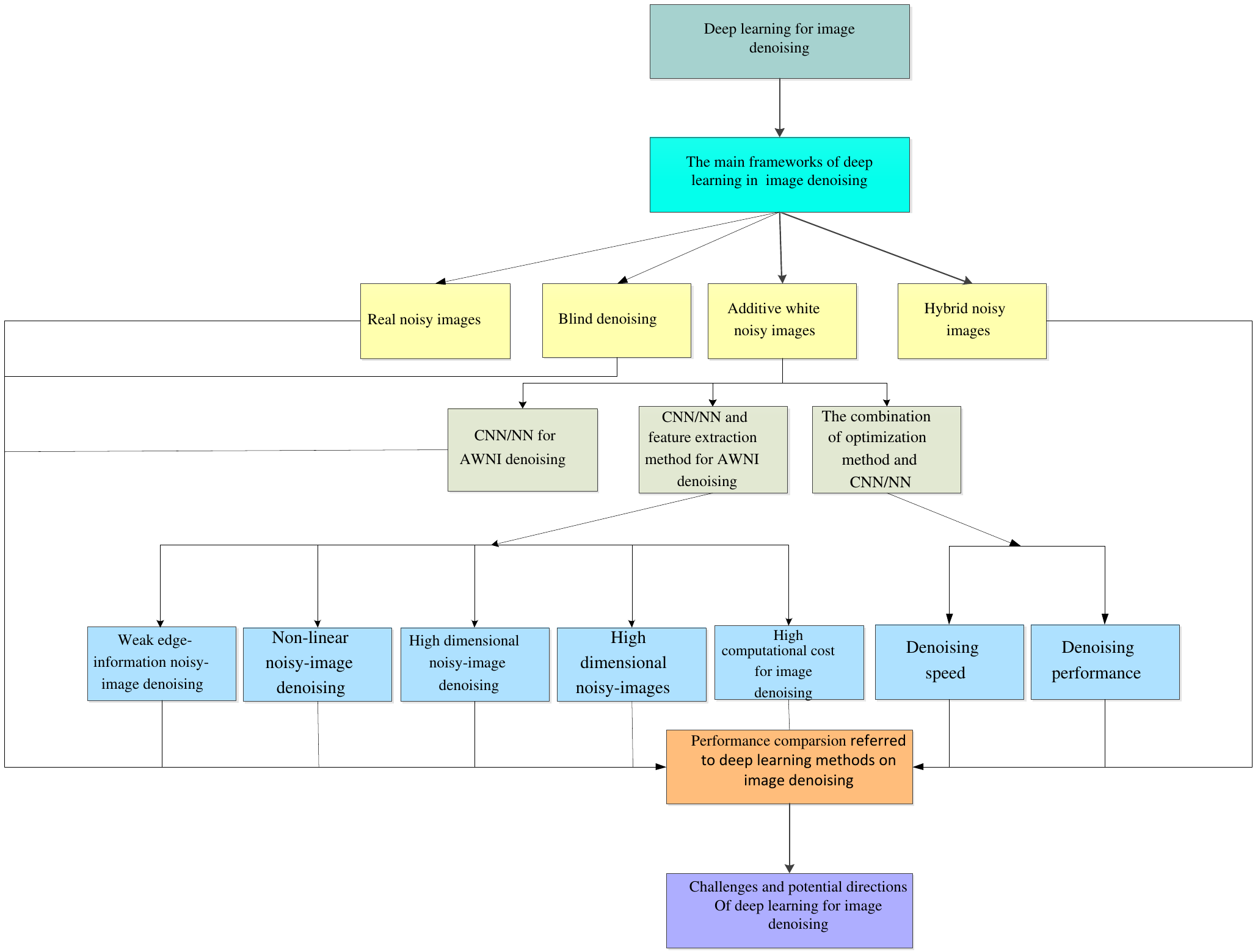}
\label{fig_second_case}}
\caption{Outline of the survey. It consists of four parts, including basic frameworks, categories, performance comparison, challenges and potential directions. Specifically, categories comprise additive white noisy images, real noisy images, blind denoising and hybrid noisy images.}
\label{fig:5}
\end{figure}

In terms of blind denoising, a fast and flexible denoising CNN (FFDNet) \cite{zhang2018ffdnet} presented different noise levels and the noisy image patch as the input of a denoising network to improve denoising speed and process blind denoising. For handling unpaired noisy images, a generative adversarial network (GAN) CNN blind denoiser (GCBD) \cite{chen2018image} resolved this problem by first generating the ground truth, then inputting the obtained ground truth into the GAN to train the denoiser. Alternatively, a convolutional blind denoising network (CBDNet) \cite{guo2019toward} removed the noise from the given real noisy image by two sub-networks, one in charge of estimating the noise of the real noisy image, and the other for obtaining the latent clean image. For more complex corrupted images, a deep plug-and-play super-resolution (DPSR) method \cite{zhang2019deep} was developed to estimate blur kernel and noise, and recover a high-resolution image. Although other important research has been conducted in the field of image denoising in recent years, there have been only a few reviews to summarize the deep learning techniques in image denoising \cite{tian2018deep}. Although Ref. \cite{tian2018deep} referred to a good deal work, it lacked more detailed classification information about deep learning for image denoising. For example, related work pretaining to unpaired real noisy images was not covered. To this end, we aim to provide an overview of deep learning for image denoising, in terms of both applications and analysis. Finally, we discuss the state-of-the-art methods for image denoising, including how they can be further expanded to respond to the challenges of the future, as well as potential research directions. An outline of this survey is shown in Fig. 1.

This overview covers more than 200 papers about deep learning for image denoising in recent years. The main contributions in this paper can be summarized as follows.

1. The overview illustrates the effects of deep learning methods on the field of image denoising.

2. The overview summarizes the solutions of deep learning techniques for different types of noise (i.e., additive white noise, blind noise, real noise and hybrid noise) and analyzes the motivations and principles of these methods in image denoising, where blind noise denotes noise of unknown types.
Finally, we evaluate the denoising performance of these methods in terms of quantitative and qualitative analysis.

3. The overview points out some potential challenges and directions for deep learning in the use of image denoising.

The rest of this overview is organized as followed.

Section 2 discusses the popular deep learning frameworks for image applications. Section 3 presents the main categories of deep learning in image denoising, as well as a comparison and analysis of these methods. Section 4 offers a performance comparison of these denoising methods. Section 5 discusses the remaining challenges and potential research directions. Section 6 offers the authors' conclusions.
\section{Fundamental frameworks of deep learning methods for image denoising}
This section offers a discussion of deep learning, including the ideas behind it, the main network frameworks (techniques), and the hardware and software, which is the basis for the deep learning techniques for image denoising covered in this survey.
\subsection{Machine learning methods for image denoising}
Machine learning methods consist of supervised, semi-supervised and unsupervised learning methods. Supervised learning methods \cite{litjens2017survey,xiao2019local,li2019shared} use the given label to put the obtained features closer to the target for learning parameters and training the denoising model. For example, take a given denoising model $y = x + \mu$, where $x$, $y$ and $\mu$ represent the given clean image, noisy image and additive Gaussian noise (AWGN) of standard deviation $\sigma$, respectively. From the equation above and Bayesian knowledge, it can be seen that the learning of parameters of the denoising model relies on pair $\{ {x_k},{y_k}\} _{k = 1}^N$, where ${x_k}$ and ${y_k}$ denote the $kth$ clean image and noisy image, respectively. Also, $N$ is the number of noisy images. This processing can be expressed as ${x_k} = f({y_k},\theta ,m)$, where $\theta$ is the parameters and $m$ denotes the given noise level.

Unsupervised learning methods \cite{lee2018unsupervised} use given training samples to find patterns rather than label matching and finish specific tasks, such as unpairing real low-resolution images \cite{yuan2018unsupervised}. The recently proposed Cycle-in-Cycle GAN (CinCGAN) recovered a high-resolution image by first estimating the high-resolution image as a label, then exploiting the obtained label and loss function to train the super-resolution model.

Semi-supervised learning methods \cite{choi2019semi} apply a model from a given data distribution to build a learner for labeling unlabeled samples. This mechanism is favored by small sample tasks, such as medical diagnosis. A semi-supervised learned sinogram restoration network (SLSR-Net) can learn feature distribution from paired sinograms via a supervised network, and then, convert the obtained feature distribution to a high-fidelity sinogram from unlabeled low-dose sinograms via an unsupervised network \cite{meng2020semi}.
\subsection{Neural networks for image denoising}
Neural networks are the basis of machine learning methods, which in turn are the basis of deep learning techniques \cite{schmidhuber2015deep}. Most neural networks consist of neurons, input $X$, activation function $f$, weights $W = [{W^0},{W^1},...,{W^{n - 1}}]$ and biases $b = [{b^0},{b^1},...,{b^n}]$. The activation functions such as Sigmoid \cite{marreiros2008population,karlik2011performance} and tanh \cite{jarrett2009best,fan2000extended} can convert the linear input into non-linearity through $W$ and $b$ as follows.
\begin{equation}\label{2}
f(X;W;b) = f({W^T}X + b).
\end{equation}

Note that if the neural network has multiple layers, it is regarded as multilayer perceptron (MLP) \cite{burger2012image}. In addition, the middle layers are treated as hidden layers beside the input and output layers. This process can be expressed as
\begin{equation}\label{2}
f(X;W;b) = f({W^n}f({W^{n - 1}}...f({W^0}X + {b^0})...{b^{n - 1}}) + {b^n}),
\end{equation}
where $n$ is the final layer of the neural network. To help readers understand the principle of the neural network, a visual example is provided in Fig. 2.

\begin{figure}[!htb]
\centering
\subfloat{\includegraphics[width=3in]{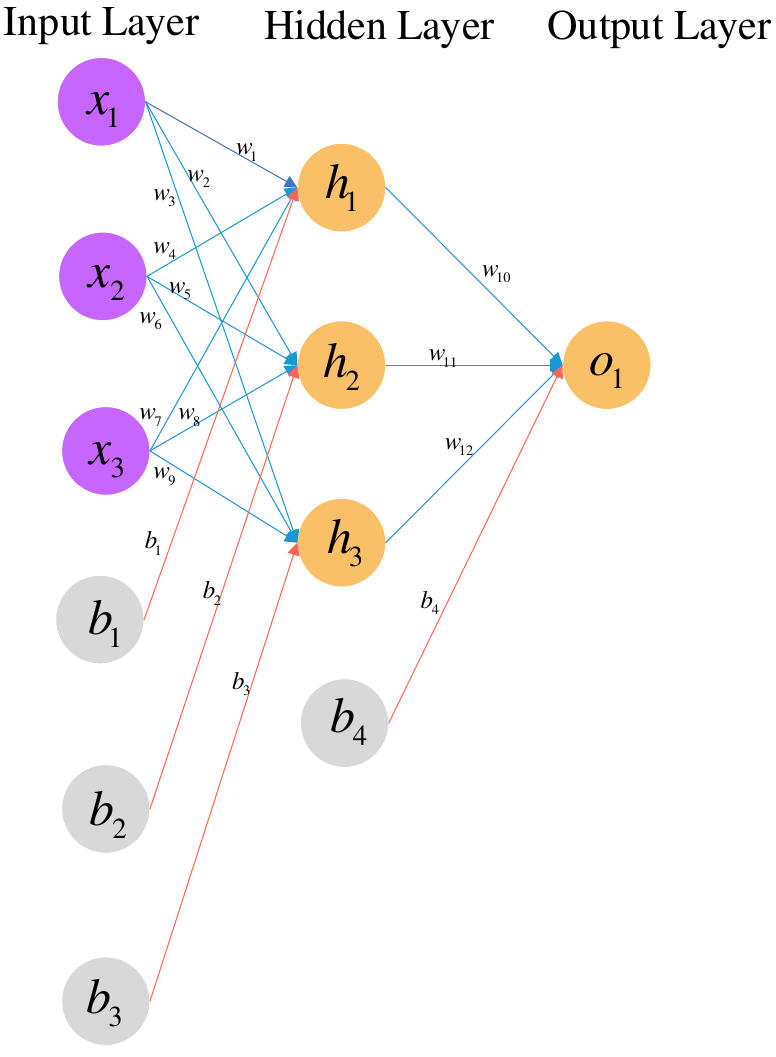}
\label{fig_second_case}}
\caption{Two-layer neural network.}
\label{fig:5}
\end{figure}

The two-layer fully connected neural network includes two layers: a hidden layer and output layer (the input layer is not generally regarded as a layer of a  neural network). There are parameters to be defined: ${x_1}$, ${x_2}$, ${x_3}$ and ${o_1}$ represent the inputs and output of this neural network, respectively. ${w_1},{w_2},...,{w_{11}},{w_{12}}$ and ${b_1}$, ${b_2}$, ${b_3}$, ${b_4}$ are the weights and biases, respectively. For example, the output of one neuron ${h_1}$ via Eqs. (3) and (4) is obtained as follows:
\begin{equation}\label{2}
f({z_{h1}}) = f({w_1}{x_1} + {w_4}{x_2} + {w_7}{x_3} + {b_1}).
\end{equation}
\begin{equation}\label{2}
o({h_1}) = f({z_{h1}}).
\end{equation}

First, the output of the network ${o_1}$ is obtained. Then, the network uses back propagation (BP) \cite{hirose1991back} and loss function to learn parameters. That is, when the loss value is within specified limitation, the trained model is considered as well-trained. It should be noted that if the number of layers of a neural network is more than three, it is also referred to as a deep neural network. Stacked auto-encoders (SARs) \cite{hinton2006reducing} and deep belief networks (DBNs) \cite{bengio2007greedy,hinton182006} are typical deep neural networks. They used stacked layers in an unsupervised manner to train the models and obtain good performance. However, these networks are not simple to implement and require a good deal of manual settings to achieve an optimal model. Due to this, end-to-end connected networks, especially CNNs, were proposed \cite{yao2018joint}. CNNs have wide applications in the field of image processing, especially image denoising.
\subsection{CNNs for image denoising}
Due to their plug-and-play network architectures, CNNs have achieved great success in image processing \cite{zhang2018ista,lu2018fast,li2017shared}. As a pioneer in CNN technology, LeNet \cite{lecun1998gradient} used convolutional kernels of different sizes to extract features and obtain good performance in image classification. However, due to the Sigmoid activation function, LeNet had a slow convergence speed, which was a shortcoming in real-world applications.

After LeNet, the proposed AlexNet \cite{krizhevsky2012imagenet} was a milestone for deep learning. Its success was due to several reasons. First, the graphics processing unit (GPU) \cite{marreiros2008population} provided strong computational ability. Second, random clipping (i.e., dropout) solved the overfitting problem. Third, ReLU \cite{nair2010rectified} improved the speed of stochastic gradient descent (SGD) rather than Sigmoid \cite{bottou2010large}. Fourth, the data augmentation method further addressed the overfitting problem. Although AlexNet achieved good performance, it required substantial memory usage due to its large convolutional kernels. That limited its real-world applications, such as in smart cameras. After that, during the period of 2014 to 2016, deeper network architectures with small filters were preferred to improve the performance and reduce computational costs. Specifically, VGG \cite{simonyan2014very} stacked more convolutions with small kernel sizes to win the ImageNet LSVR Challenge in 2014. Fig. 3 depicts the network architecture.
\begin{figure}[!htb]
\centering
\subfloat{\includegraphics[width=6.5in]{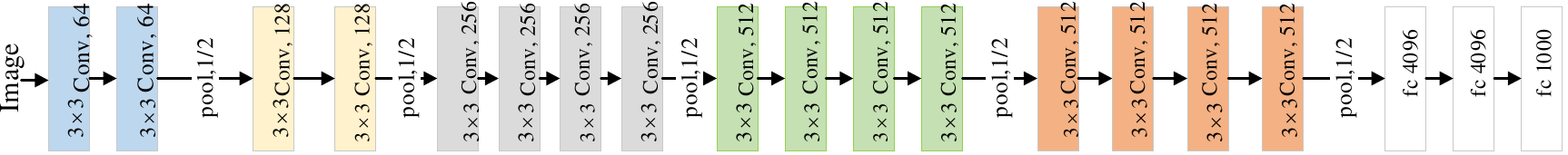}
\label{fig_second_case}}
\caption{Network architecture of VGG.}
\label{fig:5}
\end{figure}

 With the success of deeper networks, the research turned to increasing their width. GoogLeNet \cite{szegedy2015going} increased the width to improve the performance for image applications. Moreover, GoogLeNet transformed a large convolutional kernel into two smaller convolution kernels in order to reduce the number of parameters and computational cost. GoogLeNet also used the inception module \cite{lin2013network} as well as Inception 1. Its visual network figure is shown in Fig. 4.
\begin{figure}[!htb]
\centering
\subfloat{\includegraphics[width=4.5in]{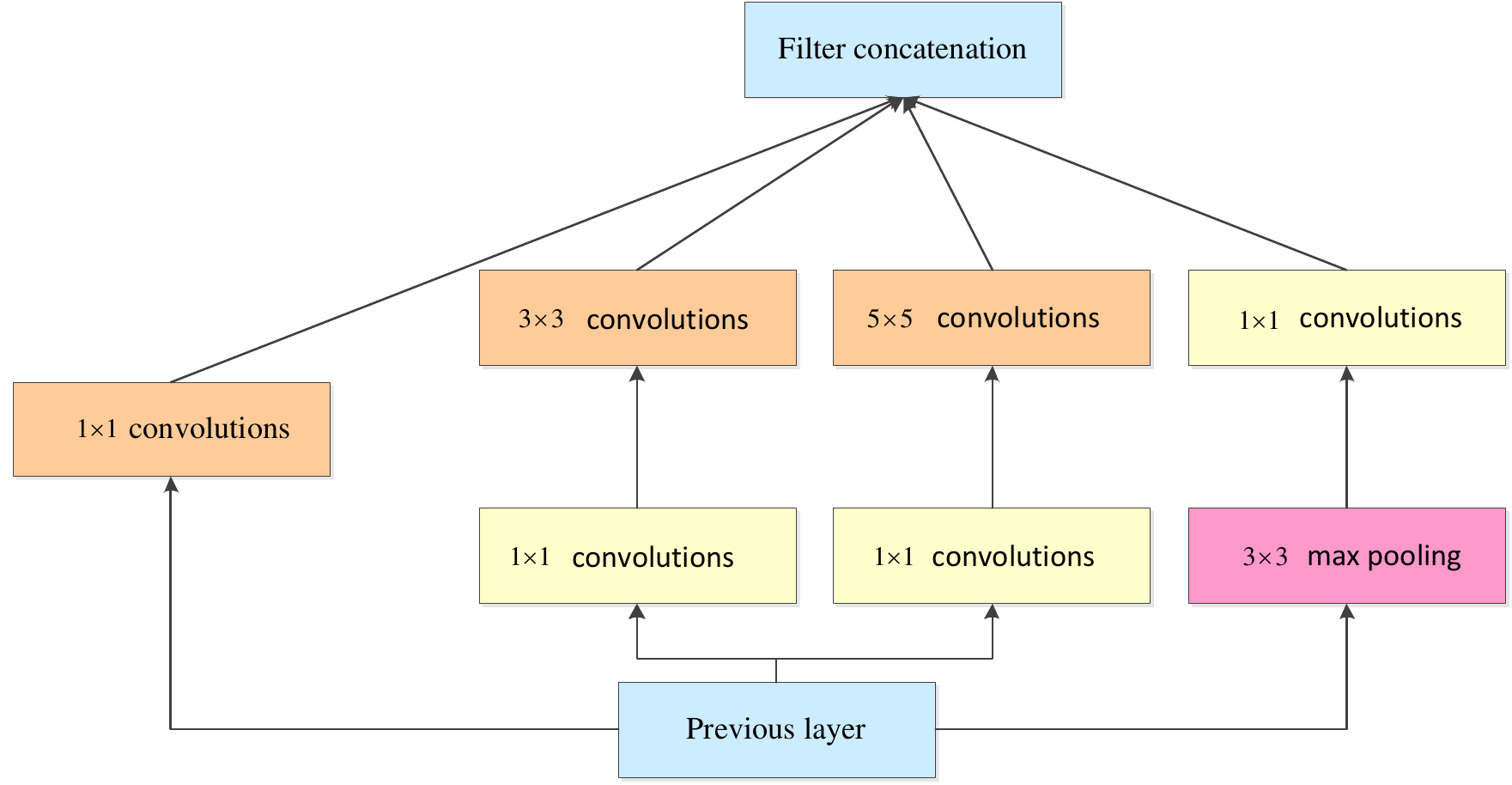}
\label{fig_second_case}}
\caption{Network architecture of GoogLeNet (Inception 1).}
\label{fig:5}
\end{figure}

Although VGG and GoogLeNet methods are effective for image applications, they have two drawbacks: if the network is very deep, this may result in vanishing or exploding gradients; and if the network is very wide, it may be subject to the phenomenon of overfitting. To overcome these problems, ResNet \cite{he2016deep} was proposed in 2016. Each block was given by adding residual learning operation in ResNet to improve the performance of image recognition, which leads to ResNet winning the mageNet LSVR in 2015. Fig. 5 depicts the concept of residual learning.
\begin{figure}[!htb]
\centering
\subfloat{\includegraphics[width=1.2in]{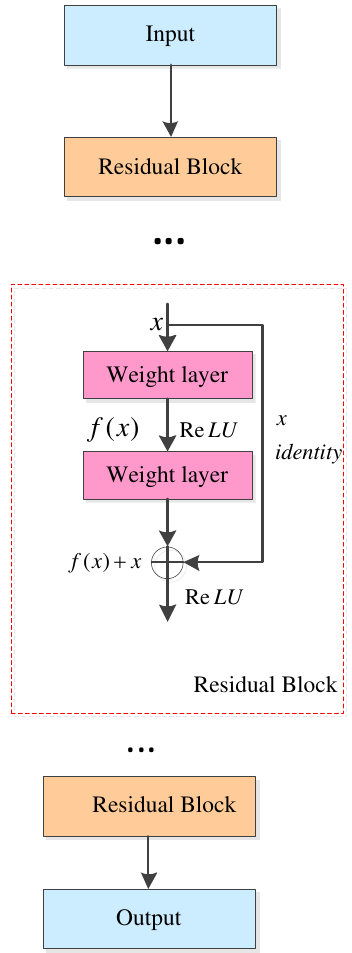}
\label{fig_second_case}}
\caption{Network architecture of ResNet.}
\label{fig:5}
\end{figure}

Since 2014, deep networks have been widely used in real-world image applications, such as facial recognition \cite{hu2015face} and medical diagnosis \cite{li2014medical}. However, in many applications, captured images, such as real noisy images, are not sufficient, and deep CNNs tend to perform poorly in image applications. For this reason, GANs \cite{radford2015unsupervised} were developed. GANs consisted of two networks: generative and discriminative networks. The generative network (also referred to as the generator) is used to generate samples, according to input samples. The discriminative network (also called the discriminator) is used to judge the truth of both input samples and generated samples. The two networks are adversarial. Note that if the discriminator can accurately distinguish real samples and generate samples from generator, the trained model is regarded as finished. The network architecture of the GAN can be seen in Fig. 6. Due to its ability to construct supplemental training samples, the GAN is very effective for small sample tasks, such as facial recognition \cite{tran2017disentangled} and complex noisy image denoising \cite{chen2018image}. These mentioned CNNs are basic networks for image denoising.
\begin{figure}[!htb]
\centering
\subfloat{\includegraphics[width=3.5in]{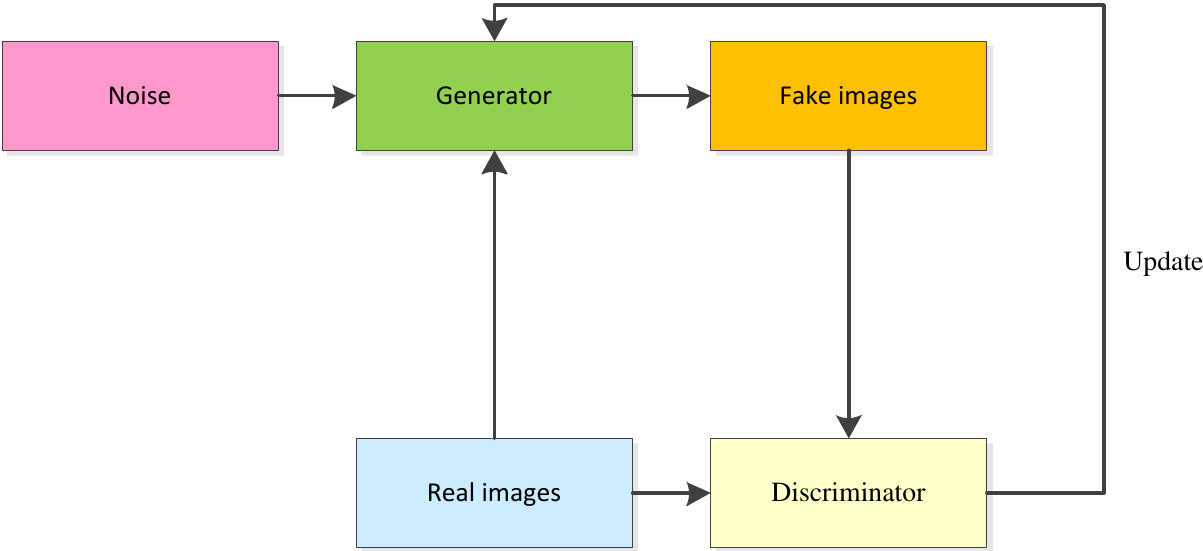}
\label{fig_second_case}}
\caption{Network architecture of GAN.}
\label{fig:5}
\end{figure}
\subsection{Hardware and software used in deep learning}
One reason for the success of deep learning is the GPU. The GPU uses the CUDA \cite{nvidia2011nvidia}, OpenCL \cite{stone2010opencl} and cuDNN \cite{chetlur2014cudnn} platforms to strengthen its parallel computing ability, which exceeds the speed of the CPU by 10 to 30 times. The GPU consists of an NVIDIA consumer line of graphics cards (i.e., GTX 680, GTX 980, GTX 1070, GTX 1070Ti, GTX1080, GTX 1080Ti, RTX 2070, RTX 2080, RTX 2080Ti, Tesla K40c, Tesla K80, Quadro M6000, Quadro GP100, Quadro P6000 and Tesla V100) and AMD (i.e., Radeon Vega 64 and FE) \cite{kutzner2019more}.

Deep learning software can provide interfaces to call the GPU. Popular software packages include:

(1)	Caffe \cite{jia2014caffe} based on C++, provides
C++, Python and Matlab interfaces, which can also run on both the CPU and GPU. It is widely used for object detection tasks. However, it requires developers to master C++.

(2) Theano \cite{bergstra2010theano} is a compiler of math expressions for dealing with large-scale neural networks. Theano provides a Python interface and is used in image super-resolution, denoising and classification.

(3) Matconvnet \cite{vedaldi2015matconvnet} offers the Matlab interface. It is utilized in image classification, denoising and super-resolution, and video tracking. However, it requires Matlab mastery.

(4) TensorFlow \cite{abadi2016tensorflow} is a relatively high-order machine learning library. It is faster than Theano for compilation. TensorFlow offers C++ and Python interfaces and is used in object detection, image classification, denoising and super-resolution.

(5) Keras \cite{chollet2015keras} based on TensorFlow and Theano is implemented in Python and offers a Python interface. It can be used in image classification, object detection, image resolution, image denoising and action recognition.

(6) PyTorch \cite{paszke2017automatic} is implemented in Python and offers a Python interface. It is employed in image classification, object detection, image segmentation, action recognition, image super-resolution, image denoising and video tracking.
\section{Deep learning techniques in image denoising}
\subsection{Deep learning techniques for additive white noisy-image denoising}
Due to the insufficiency of real noisy images, additive white noisy images (AWNIs) are widely used to train the denoising model \cite{jin2017deep}. AWNIs include Gaussian, Poisson, Salt, Pepper and multiplicative noisy images \cite{farooque2013survey}. There are several deep learning techniques for AWNI denoising, including CNN/NN; the combination of CNN/NN and common feature extraction methods; and the combination of the optimization method and CNN/NN.
\subsubsection{CNN/NN for AWNI denoising}
Automatic feature extraction methods can play a major role in reducing the computational costs for image applications \cite{yang2012beyond,ren2019low,lu2018low}. For this reason, CNNs have been developed for image denoising \cite{mccann2017convolutional,liu2017image}. Zhang et al. \cite{zhang2017beyond} proposed a model as well as a DnCNN to deal with multiple low-level vision tasks, i.e., image denoising, super-resolution and deblocking through CNN, batch normalization \cite{ioffe2015batch} and residual learning techniques \cite{he2016deep}. Wang et al. \cite{wang2017dilated}, Bae et al. \cite{bae2017beyond} and Jifara et al. \cite{jifara2019medical} also presented a residual learning into deeper CNN for image denoising. However, the deeper CNN technique relied on a deeper layer rather than a shallow layer, which resulted in a long-term dependency problem. Several signal-base methods were proposed to resolve this problem. Tai et al. \cite{tai2017memnet} exploited recursive and gate units to adaptively mine more accurate features and recover clean images. Inspired by a low-rank Hankel matrix in low-level vision, Ye et al. \cite{ye2018deep} provided convolution frames to explain the connection between signal processing and deep learning by convolving local and nonlocal bases. For solving insufficient noisy images (i.e., hyperspectral and medical images), several recent works have attempted to extract more useful information through the use of improved CNNs \cite{chang2018hsi,heinrich2018residual,yu2018generative,liu20193}. For example, Yuan et al. \cite{yuan2018hyperspectral} combined a deep CNN, residual learning and multiscale knowledge to remove the noise from hyperspectral-noisy images. However, these proposed CNNs led to the likelihood of increased computational costs and memory consumption, which was not conducive for real-world applications. To address this phenomenon, Gholizadeh et al. \cite{gholizadeh2018low} utilized dilated convolutions \cite{gashi2017multi} to enlarge the receptive field and reduce the depth of network without incurring extra costs for CT image denoising. Lian et al. \cite{su2019multi} proposed a residual network via multi-scale cross-path concatenation to suppress the noise. Most of the above methods relied on improved CNNs to deal with the noise. Therefore, designing network architectures is important for image denoising \cite{park2018analysis,li2018deep}.
\begin{table}[t!]
\caption{CNN/NN for AWNI denoising.}
\label{tab:1}
\centering
\scalebox{0.44}[0.50]{
\begin{tabular}{|c|c|c|c|}
\hline
References &Methods &Applications &Key words (remarks)\\	
\hline
Zhang et al. (2017) \cite{zhang2017beyond}	&CNN	&Gaussian image denoising, super-resolution and JPEG deblocking	&CNN with residual learning, and BN for image denoising\\
\hline
Wang et al. (2017) \cite{wang2017dilated}	&CNN	&Gaussian image denoising	&CNN with dilated convolutions, and BN for image denoising\\
\hline
Bae et al. (2017) \cite{bae2017beyond}	&CNN	&Gaussian image denoisng, super-resolution	&CNN with wavelet domain, and residual learning (RL) for image restoration\\
\hline
Jin et al. (2017) \cite{jifara2019medical}	&CNN	&Medical (X-ray) image restoration	&Improved Unet from iterative shrinkage idea for medical image restoration\\
\hline
Tai et al. (2017) \cite{tai2017memnet}	&CNN	&Gaussian image denoisng, super-resolution and JPEG deblocking	&CNN with recursive unit, gate unit for image restoration\\
\hline
Anwar et al. (2017) \cite{anwar2017chaining}	&CNN	&Gaussian image denoisng	&CNN with fully connected layer, RL and dilated convolutions for image denoising\\
\hline
McCann et al. (2017)] \cite{jin2017deep} 	&CNN	&Inverse problems (i.e., denoising, deconvolution, super-resolution)	&CNN for inverse problems\\
\hline
Ye et al. (2018) \cite{ye2018deep}	&CNN	&Inverse problems(i.e., Gaussian image denoising, super-resoluion)	&Signal processing ideas guide CNN for inverse problems\\
\hline
Yuan et al. (2018) \cite{yuan2018hyperspectral}	&CNN	&Hyper-spectral image denoising	 &CNN with multiscale, multilevel features techniques for hyper-spectral image denoising\\
\hline
Jiang et al. (2018) \cite{jiang2018denoising} &CNN	&Gaussian image denoising	&Multi-channel CNN for image denoising\\
\hline
Chang et al. (2018) \cite{chang2018hsi} 	&CNN	&Hyper-spectral image (HSI) denoising, HIS restoration 	&CNN consolidated spectral and spatial coins for hyper-spectral image denoising\\
\hline
Jeon et al. (2018) \cite{jeon2018speckle} 	&CNN	&Speckle noise reduction from digital holographic images	&Speckle noise reduction of digital holographic image from Multi-scale CNN \\
\hline
Gholizadeh-Ansari et al. (2018) \cite{gholizadeh2018low}	&CNN	&Low-dose CT image denoising, X-ray image denosing	&CNN with dilated convolutions for low-dose CT image denoising\\
\hline
Uchida et al. (2018) \cite{uchida2018non}  	&CNN	&Non-blind image denoising	&CNN with residual learning for non-blind image denoising\\
\hline
Xiao et al. (2018) \cite{xiao2018removing}	&CNN	&Stripe noise reduction of infrared cloud images	&CNN with skip connection for infrared-cloud-image denoising\\
\hline
Chen et al. (2018)  \cite{chen2018imaged}	&CNN	&Gaussian image denoisng, blind denoising	&CNN based on RL and perceptual loss for edge enhancement\\
\hline
Yu et al. (2018) \cite{yu2019deep}		&CNN	&Seismic, random, linear and multiple noise reduction of images	&A survey on deep learning for three applications\\
\hline
Yu et al. (2018) \cite{yu2018generative}	&CNN	&Optical coherence tomography (OCT) image denoising	&GAN with dense skip connection for optical coherence tomography image denoising\\
\hline
Li et al. (2018) \cite{li2018deep}	&CNN	&Ground-roll noise reduction	&An overview of deep learning techniques on ground-roll noise\\
\hline
Abbasi et al. (2018) \cite{abbasi2019three}	&CNN	&OCT image denoising	&Fully CNN with multiple inputs, and RL for OCT image denoising\\
\hline
Zarshenas et al. (2018) \cite{zarshenas2018deep} 	&CNN	&Gaussian noisy image denoising	&Deep CNN with internal and external residual learning for image denoising\\
\hline
Chen et al. (2018) \cite{chen2018deep}	&CNN 	&Gaussian noisy image denoising	&CNN with recursive operations for image denoising\\
\hline
Panda et al. (2018) \cite{panda2018exponential}	&CNN	&Gaussian noisy image denoising	&CNN with exponential linear units, and dilated convolutions for image denoising\\
\hline
Sheremet et al. (2018) \cite{sheremet2018convolutional}	&CNN &Image denoising from info-communication systems	&CNN on image denoising from info-communication systems\\
\hline
Chen et al. (2018) \cite{chen2018aerial}	&CNN	&Aerial-image denoising	&CNN with multi-scale technique, and RL for aerial-image denoising\\
\hline
Pardasani et al. (2018) \cite{pardasani2018image}	&CNN	&Gaussian, poisson or any additive-white noise reduction &CNN with BN for image denoising\\
\hline
Couturier et al. (2018) \cite{couturier2018image}	&NN	&Gaussian and multiplicative speckle noise reduction	&Encoder-decoder network with multiple skip connections for image denoising\\
\hline
Park et al. (2018) \cite{park2018analysis}	&CNN	&Gaussian noisy image denoising 	&CNN with dilated convolutions for image denoising\\
\hline
Priyanka et al. (2019) \cite{priyanka2019fully}	&CNN &Gaussian noisy image denoisng	 &CNN with symmetric network architecture for image denoising\\
\hline
Lian et al. (2019) \cite{su2019multi}	&CNN	&Poisson-noise-image denoising	&CNN with multi scale, and multiple skip connections for Poisson image denoising\\
\hline
Tripathi et al. (2018) \cite{tripathi2018correction}	&CNN	&Gaussian noisy image denoising	&GAN for image denoising\\
\hline
Zheng et al. (2019) \cite{zheng2019denoising}      	&CNN	&Gaussian noisy image denoising   &CNN for image denoising\\
\hline
Tian et al. (2019)  \cite{tian2019enhanced}     &CNN &Gaussian noisy image denoising   &CNN for image denoising\\
\hline
Remez et al. (2018) \cite{remez2018class} &CNN	&Gaussian and Poisson image denoising   &CNN for image denoising\\	
\hline
Tian et al. (2020) \cite{tian2020image}	&CNN	&Gaussian image denoising and real noisy image denoising	&CNN with BRN, RL and dilated convolutions for image denosing\\
\hline
Tian et al. (2020) \cite{tian2020attention} &CNN &Gaussian image denoising, blind denoising and real noisy image denoising &CNN with attention mechanism and sparse method for image denoising\\
\hline
Tian et al. (2020) \cite{tian2020designing} &CNN  &Gaussian image denoising, blind denoising and real noisy image denoising &Two CNNs with sparse method for image denoising\\
\hline
\end{tabular}}
\label{tab:booktabs}
\end{table}\

Changing network architectures involves the following methods \cite{yu2019deep,mafi2018comprehensive,ren2019progressive}: fusing features from multiple inputs of a CNN; changing the loss function; increasing depth or width of the CNN; adding some auxiliary plug-ins into CNNs; and introducing skip connections or cascade operations into CNNs. Specifically, the first method includes three types: different parts of one sample as multiple inputs from different networks \cite{abbasi2019three}; different perspectives for the one sample as input, such as multiple scales \cite{jeon2018speckle,chen2018aerial}; and different channels of a CNN as input \cite{jiang2018denoising}. The second method involves the design of different loss functions according to the characteristics of nature images to extract more robust features \cite{aljadaany2019proximal}. For example, Chen et al. \cite{chen2018imaged} jointed Euclidean and perceptual loss functions to mine more edge information for image denoising. The third method enlarged the receptive field size to improve denoising performance via increasing the depth or width of the network \cite{uchida2018non,zarshenas2018deep,sheremet2018convolutional}. The fourth method applied plug-ins, such as activation function, dilated convolution, fully connected layer and pooling operations, to enhance the expressive ability of the CNN \cite{panda2018exponential,priyanka2019fully,pardasani2018image}. The fifth method utilized skip connections \cite{xiao2018removing,chen2018deep,couturier2018image,anwar2017chaining} or cascade operations \cite{su2019multi,chen2019end} to provide complementary information for the deep layer in a CNN. Table 1 provides an overview of CNNs for AWNI denoising.
\subsubsection{CNN/NN and common feature extraction methods for AWNI denoising}
Feature extraction is used to represent the entire image in image processing, and it is important for machine learning \cite{liang2019novel,lu2019robust,yang2013sparse}. However, because deep learning techniques are black box techniques, they do not allow the choice of choose features, and therefore cannot guarantee that the obtained features are the most robust \cite{shwartz2017opening,wei2019m3net}. Motivated by problem, researches embedded common feature extraction methods into CNNs for the purpose of image denoising. They did this for five reasons: weak edge-information noisy images, non-linear noisy images, high dimensional noisy images and non-salient noisy images, and high computational costs.

For weak edge-information noisy images, CNN with transformation domain methods were proposed by Guan et al. \cite{guan2019wavelet}, Li et al. \cite{li2018cnn}, Liu et al. \cite{liu2018multi}, Latif et al. \cite{latif2018deep} and Yang et al. \cite{yang2017bm3d}. However, they were not effective in removing the noise.  Specifically, in \cite{liu2018multi}, the proposed solution used the wavelet method and U-net to eliminate the gridding effect of dilated convolutions on enlarging the receptive field for image restoration.

For non-linear noisy images, CNNs with kernel methods proved useful \cite{bako2017kernel,xu2019learning}. These methods mostly consisted of three steps \cite{mildenhall2018burst}. The first step used CNN to extract features. The second step utilized the kernel method to convert obtained non-linear features into linearity. The third step exploited the residual learning to construct the latent clean image.

For high dimensional noisy images, the combination of CNN and the dimensional reduction method was proposed \cite{xie2018deep,guo2018deep}. For example, Khaw et al. \cite{khaw2017image} used a CNN with principal component analysis (PCA) for image denoising. This consisted of three steps. The first step used convolution operations to extract features. The second step utilized the PCA to reduce the dimension of the obtained features. The third step employed convolutions to deal with the obtained features from the PCA and to reconstruct a clean image.

For non-salient noisy images, signal processing can guide the CNN in extracting salient features \cite{jia2018multiscale,kadimesetty2018convolutional,ran2019denoising,abbasi2019three}. Specifically, skip connection is a typical operation of signal processing \cite{kadimesetty2018convolutional}.
\begin{table}[H]
\caption{CNN/NN and common feature extraction methods for AWNI denoising.}
\label{tab:1}
\centering
\scalebox{0.50}[0.50]{
\begin{tabular}{|c|c|c|c|}
\hline
References &Methods &Applications &Key words (remarks)\\	
\hline
Bako et al. (2017) \cite{su2019multi}	&CNN	&Monte Carlo-rendered images denoising	&CNN with kernel method for estimating noise piexls\\	
\hline
Ahn et al. (2017) \cite{ahn2017block}	&CNN	&Gaussian image denoising 	&CNN with NSS for image denoising\\
\hline
Khaw et al. (2017) \cite{khaw2017image}	&CNN	&Impulse noise reduction	&CNN with PCA for image denoising\\
\hline
Vogel et al. (2017) \cite{vogel2017primal}	&CNN	&Gaussian image denoising	&U-net with multi scales technique for image denoising\\
\hline
Mildenhall et al. (2018) \cite{mildenhall2018burst}	&NN	&Low-light synthetic noisy image denoising, real noise	&Encoder-decoder with multi scales, and kernel method for image denoising\\
\hline
Liu et al. (2018) \cite{liu2018multi}	&CNN	&Gaussian image denoisng, super-resolution and JPEG deblocking 	&U-net with wavelet for image restoration\\
\hline
Yang et al. (2018) \cite{yang2017bm3d}	&CNN &Gaussian image denoisng	&CNN with BM3D for image denoising\\
\hline
Guo et al. (2018) \cite{guo2018deep}	&CNN	&Image blurring and denoising	&CNN with RL, and sparse method for image denoising\\
\hline
Jia et al. (2018) \cite{jia2018multiscale}	&CNN	&Gaussian image denoisng	&CNN with multi scales, and dense RL operations for image denoising\\
\hline
Ran et al. (2018) \cite{ran2019denoising}	&CNN	&OCT image denoising, OCT image super-resolution	&CNN with multi views for image restoration\\
\hline
Li et al. (2018) [140]	&CNN	&Medical image denoising, stomach pathological image denoising	&CNN consolidated wavelet for medical image denoising\\
\hline
Ahn et al. (2018) \cite{ahn2018block}	&CNN &Gaussian image denoisng	&CNN with NSS for image denoising\\
\hline
Xie et al. (2018) \cite{xie2018deep}	&CNN &Hyper-spectral image denoising	&CNN with RL, and PCA for low-dose CT image denoising\\
\hline
Kadimesetty et al. (2019) \cite{kadimesetty2018convolutional}	&CNN &Low-Dose computed tomography (CT) image denoising	&CNN with RL, batch normalization (BN) for medical image denoising\\
\hline
Guan et al. (2019) \cite{guan2019wavelet}	&CNN &Stripe noise reduction 	&CNN with wavelet-image denoising\\
\hline
Abbasi et al. (2019) \cite{abbasi2019three}	&NN	&3D magnetic resonance image denoising, medical image denoising	&GAN based on encoder-decoder and RL for medical denoising\\
\hline
Xu et al. (2019) \cite{xu2019learning}	&CNN &Synthetic and real noisy and video denoising	&CNN based on deformable kernel for image and video denoising\\
\hline
\end{tabular}}
\label{tab:booktabs}
\end{table}\

For tasks involving high computational costs, a CNN with relations nature of pixels from an image was very effective in decreasing complexity \cite{abbasi2019three,ahn2018block,ahn2017block}. For example, Ahn et al. \cite{ahn2017block} used a CNN with non-local self-similarity (NSS) to filter the noise, where similar characteristics of the given noisy image can accelerate the speed of extraction feature and reduce computational costs.

More detailed information on these methods mentioned can be found in Table 2.
\subsubsection{Combination of optimization method and CNN/NN for AWNI denoising}
Machine learning uses optimization techniques \cite{hsu2017cnn,li2020robust} and discriminative learning methods \cite{li2019fast,liu2017wide} to deal with image applications. Although optimization methods have good performance on different low-level vision tasks, these methods need manual setting parameters, which are time-consuming \cite{tian2020coarse,tian2020lightweight}. The discriminative learning methods are fast in image restoration. However, they are not flexible for low-level vision tasks. To achieve a tradeoff between efficiency and flexibility, a discriminative learning optimization-based method \cite{meinhardt2017learning,bigdeli2017image} was presented for image applications, such as image denoising. CNNs with prior knowledge via regular term of loss function is a common method in image denoising \cite{hongqiang2018adaptive}, which can be divided two categories: improvement of denoising speed and improvement of denoising performance.

For improving denoising speed, an optimization method using a CNN was an effective tool for rapidly finding the optimal solution in image denoising \cite{cho2018gradient,fu2019convolutional}. For example, a GAN with the maximum a posteriori (MAP) method was used to estimate the noise and deal with other tasks, such as image inpainting and super-resolution \cite{yeh2018image}. An experience-based greed algorithm and transfer learning strategies with a CNN can accelerate a genetic algorithm to obtain a clean image \cite{liu2018neural}. Noisy image and noise level mapping were inputs of the CNN, which had faster execution in predicting the noise \cite{tassano2019analysis}.
\begin{table}[H]
\caption{The combination of the optimization method and CNN/NN for AWNI denoising.}
\label{tab:1}
\centering
\scalebox{0.48}[0.50]{
\begin{tabular}{|c|c|c|c|}
\hline
References &Methods &Applications &Key words (remarks)\\	
\hline
Hong et al. (2018) \cite{hongqiang2018adaptive}	&CNN	&Gaussian image denoising	&Auto-Encoder with BN, and ReLU for image denoising\\	
\hline
Cho et al. (2018) \cite{cho2018gradient}	&CNN	&Gaussian image denoising 	&CNN with separable convolution, and gradient prior for image denoising\\
\hline
Fu et al. (2018) \cite{fu2019convolutional}	&CNN &Salt and pepper noise removal	&CNN with non-local switching filter for salt and pepper noise\\
\hline
Yeh et al. (2018) \cite{yeh2018image}	&CNN	&Image denoising super-resolution and inpainting	&GAN with MAP for image restoration\\
\hline
Liu et al. (2018) \cite{liu2018neural} &CNN	&Medical image denoising, computed tomography perfusion for image denoising	&CNN with genetic algorithm for medical image denoising\\
\hline
Tassano et al. (2019) \cite{tassano2019analysis}	&CNN	&Gaussian image denoisng 	&CNN with noise level, upscaling, downscaling operation for image denoising\\
\hline
Heckel et al. (2018) \cite{heckel2018rate} 	&CNN	&Image denoisng	&CNN with deep prior for image denoising\\
\hline
Jiao et al. (2017) \cite{jiao2017formresnet}	&CNN	&Gaussian image denoisng, image inpainting	 &CNN with inference, residual operation for image restoration\\
\hline
Wang et al. (2017) \cite{wang2017elu}	&CNN	&Image denoising	&CNN with total variation for image denoising\\
\hline
Li  et al. (2019) \cite{li2019learning} 	&CNN	&Image painting	&CNN with split Bregman iteration algorithm for image painting\\
\hline
Sun et al. (2018) \cite{sun2018adversarial}	&CNN &Gaussian image denoisng	&GAN with skip-connections, and ResNet blocks for image denoising\\
\hline
Zhi et al. (2018) \cite{zhiping2018new}	&CNN &Gaussian image denoising	&GAN with multiscale for image denoising\\
\hline
Du et al. (2018) \cite{du2018image} 	&CNN	&Gaussian image denoising	&CNN with wavelet for medcial image restoration\\
\hline
Liu et al. (2019) \cite{liu2019dual}	&CNN &Gaussian image denoising, real noisy image denoising, rain removal &Dual CNN with residual operations for  image restoration\\
\hline
Khan et al.  (2019) \cite{khan2019symbol}	&CNN &Symbol denoising	&CNN with quadrature amplitude modulation for symbol denoising\\
\hline
Zhang et al. (2019) \cite{zhang2019vst}	&CNN &Image Possian denoising	&CNN with variance-stabilizing transformation for poisson denoising\\
\hline
Cruz et al. (2018) \cite{cruz2018nonlocality}	&CNN &Gaussian image denoising	&CNN with nonlocal filter for image denoising\\
\hline
Jia et al. (2019) \cite{jia2019focnet}	&CNN &Gaussian image denoising	&CNN based on a fractional-order differential equation for image denoising\\
\hline
\end{tabular}}
\label{tab:booktabs}
\end{table}\
For improving denoising performance, a CNN combined optimization method was used to make a noisy image smooth \cite{heckel2018rate,gondara2017recovering,jiao2017formresnet}. A CNN with total variation denoising reduced the effect of noise pixels \cite{wang2017elu}. Combining the Split Bregman iteration algorithm and CNN \cite{li2019learning} can enhance pixels through image depth to obtain a latent clean image. A dual-stage CNN with feature matching can better recover the detailed information of the clean image, especially noisy images \cite{sun2018adversarial}. The GAN with the nearest neighbor algorithm was effective in filtering out noisy images from clean images \cite{zhiping2018new}. A combined CNN used wavefront coding to enhance the pixels of latent clean images via the transform domain \cite{du2018image}. Other effective denoising methods are shown in \cite{liu2019dual,khan2019symbol,gong2018learning}. Table 3 shows detailed information about the combination of the optimization methods and CNN/NN in AWNI denoising.
\subsection{Deep learning techniques for real noisy image denoising}
There are mainly two types of deep learning techniques for image denoising: single end-to-end CNN and the combination of prior knowledge and CNN.

For the first method, changing the network architecture is an effective way to remove the noise from the given real corrupted image. Multiscale knowledge is effective for image denoising. For example, a CNN consisting of convolution, ReLU and RL employed different phase features to enhance the expressive ability of the low-light image denoising model \cite{tao2017llcnn}. To overcome the blurry and false image artifacts, a dual U-Net with skip connection was proposed for computed tomography (CT) image reconstruction \cite{han2018framing}. To address the problem of resource-constraints problem, Tian et al. \cite{tian2020image} used a dual CNN with batch renormalization \cite{ioffe2017batch}, RL and dilated convolutions to deal with real noisy images. Based on nature of light images,
two CNNs utilized anisotropic parallax analysis to generate structural parallax information for real noisy images \cite{chen2018light}. Using a CNN to resolve remote sensing \cite{jian2018low} and medical images \cite{khoroushadi2018enhancement} under low-light conditions proved effective \cite{jiang2018deep}. To extract more detailed information, recurrent connections were used to enhance the representative ability to deal with corrupted images in the real world \cite{godard2018deep,zhao2019end}. To deal with unknown real noisy images, a residual structure was utilized to facilitate low-frequency features, and then, an attention mechanism \cite{tian2020attention} could be applied to extract more potential features from channels \cite{anwar2019real}. To produce the noisy image,
a technique used imitating cameral pipelines to construct the degradation model in order to filter the real noisy images \cite{jaroensri2019generating}. To tackle the problem of unpairing noisy images, an unsupervised learning method embedded into the CNN proved effective in image denoising \cite{cui2019pet}. The self-consistent GAN \cite{yan2019unsupervised} first used a CNN to estimate the noise of the given noisy image as a label, and then, applied another CNN and the obtained label to remove the noise for other noisy images. This concept has also been extended to general CNNs. The Noise2Inverse method used a CNN to predict the value of a noisy pixel, according to
its surrounding noisy pixels \cite{hendriksen2020noise2inverse}. The attention mechanism merged into a 3D self-supervised network can improve the efficiency of removing the noise from medical noisy images \cite{li2020sacnn}.
More detailed information about the above research is shown in Table 4.
\begin{table}[H]
\caption{CNNs for real noisy image denoising.}
\label{tab:1}
\centering
\scalebox{0.47}[0.55]{
\begin{tabular}{|c|c|c|c|}
\hline
References &Methods &Applications &Key words (remarks)\\	
\hline
Tao et al. (2019) \cite{tao2017llcnn}	&CNN	&Real noisy image denoising, low-light image enhancement 	&CNN with ReLU, and RL for real noisy  image denoising\\	
\hline
Chen et al. (2018) \cite{chen2018image}   &CNN    &Real noisy image denoising, blind denoising	&GAN for real noisy image denoising\\
\hline
Han et al. (2018) \cite{han2018framing}	&CNN	&CT image reconstruction	&U-Net with skip connection for CT image reconstruction\\
\hline
Chen et al. (2018) \cite{chen2018light}	&CNN	&Real noisy image denoising 	&CNNs with anisotropic parallax analysis for real noisy image denoising\\
\hline
Jian et al. (2018) \cite{jian2018low}	&CNN	&Low-light remote sense image denoising	&CNN for image denoising\\
\hline
Khoroushadi et al. (2019) \cite{khoroushadi2018enhancement}	&CNN &Medical image denoising, CT image denoising	&CNN for image denoising\\
\hline
Jiang et al. (2018) \cite{jiang2018deep}	&CNN	 &Low-light image enhancement	&CNN with symmetric pathways for low-light image enhancement\\
\hline
Godard et al. (2018) \cite{godard2018deep}	&CNN	&Real noisy image denoising	 &CNN with recurrent connections for real noisy image denoising\\
\hline
Zhao et al. (2019) \cite{zhao2019end} 	&CNN	&Real noisy image denoising	&CNN with recurrent conncetions for real noisy image denoising\\
\hline
Anwar et al. (2019) \cite{anwar2019real} 	&CNN	&Real noisy image denoising	&CNN with RL, attention mechanism for real noisy image denoising\\
\hline
Jaroensri et al. (2019) \cite{wang2019near}	&CNN	&Real noisy image denoising	&CNN for real noisy image denoising\\
\hline
Green et al. (2018) \cite{green2018learning}    &CNN     &CT image denoising, real noisy image denoising  &CNN for real noisy image denoising\\
\hline
Brooks et al. (2019) \cite{brooks2019unprocessing}    &CNN    &Real noisy image denoising   &CNN with image processing pipeline for real noisy image denoising\\
\hline
Tian et al. (2020) \cite{tian2020image}	&CNN	&Gaussian image denoising and real noisy image denoising	&CNN with BRN, RL and dilated convolutions for image denosing\\
\hline
Tian et al. (2020) \cite{tian2020attention} &CNN &Gaussian image denoising, blind denoising and real noisy image denoising &CNN with attention mechanism and sparse method for image denoising\\
\hline
Tian et al. (2020) \cite{tian2020designing} &CNN  &Gaussian image denoising, blind denoising and real noisy image denoising &Two CNNs with sparse method for image denoising\\
\hline
Cui et al. (2019) \cite{cui2019pet} &CNN &Positron emission tomography image denoising, real noisy image denoising &Unsupervised CNN for unpair real noisy image denoising\\
\hline
Yan et al. (2019) \cite{yan2019unsupervised} &CNN &Real noisy image denoising &Self-supervised GAN for unpair real noisy image denoising\\
\hline
Broaddus et al. (2020) \cite{broaddus2020removing} &CNN &Blind denoising and real noisy image denoising &Self-supervised CNN for unpair fluorescence microscopy image denoising\\
\hline
Li et al. (2020) \cite{li2020sacnn} &CNN &CT noisy image denoising &Self-supervised CNN with attention mechanism for unpair CT image denoising\\
\hline
Hendriksen et al. (2020) \cite{hendriksen2020noise2inverse} &CNN &CT noisy image denoising &Self-supervised CNN for unpair CT image denoising\\
\hline
Wu et al. (2020) \cite{wu2020self} &CNN &CT noisy image denoising &Self-supervised CNN for unpair dynamic CT image denoising\\
\hline
\end{tabular}}
\label{tab:booktabs}
\end{table}\

The method combining CNN and prior knowledge can better deal with both speed and complex noise task in real noisy images. Zhang et al. \cite{zhang2017learning} proposed using half quadratic splitting (HQS) and CNN to estimate the noise from the given real noisy image. Guo et al. \cite{guo2019toward} proposed a three-phase denoising method. The first phase used a Gaussian noise and in-camera processing pipeline to synthesize noisy images. The synthetic and real noisy images were merged to better represent real noisy images. The second phase used a sub-network with asymmetric and total variation losses to estimate the noise of real noisy image. The third phase exploited the original noisy image and estimated noise to recover the latent clean image. To address the problem of unpaired noisy images, the combination of CNN and prior knowledge in a semi-supervised way was developed \cite{meng2020semi}. A hierarchical deep GAN (HD-GAN) first used a cluster algorithm to classify multiple categories of each patient's CT, then built a dataset by collecting the images in the same categories from different patients. Finally, the GAN was used to deal with the obtained dataset for image denoising and classification \cite{choi2019semi}. A similar method performed well in 3D mapping \cite{shantia2015indoor}.

A CNN with channel prior knowledge was effective for low-light image enhancement \cite{tao2017low}. Table 5 shows the detailed information about the above research.
\begin{table}[H]
\caption{CNNs for real noisy image denoising.}
\label{tab:1}
\centering
\scalebox{0.55}[0.60]{
\begin{tabular}{|c|c|c|c|}
\hline
References &Methods &Applications &Key words (remarks)\\	
\hline
Zhang et al. (2017) \cite{zhang2017learning}	&CNN &Real-noisy image denoising 	&CNN with HQS for real noisy image\\	
\hline
Guo et al. (2019) \cite{guo2019toward}	&CNN &Real-noisy image denoising	&CNN and cameral processing pipeline for real noisy image\\
\hline
Tao et al. (2019) \cite{tao2017low}	&CNN &Low-light image enhancement	&CNN with channel prior for low-light image enhancement\\
\hline
Ma et al. (2018) \cite{ma2018speckle}	&CNN &Tomography image denoising     &GAN with edge-prior for CT image denoising\\
\hline
Yue et al. (2018) \cite{yue2019variational} &CNN &Real-noisy image denoising, blind denoising  &CNN with variational inference for blind denoising and real-noisy image denoisng\\
\hline
Song et al. (2019) \cite{song2019dynamic} &CNN &Real noisy image denoising  &CNN with dynamic residual dense block for real noisy image denoising\\
\hline
Lin et al. (2019) \cite{lin2019real}	&CNN	&Real noisy image denoising      &GAN with attentive mechanism and noise domain for real noisy image denoising\\
\hline
Meng et al. (2020) \cite{meng2020semi} &CNN &Real noisy image denoising &CNN with semi-supervised learning for medical noisy image denoising\\
\hline
Shantia et al. (2015) \cite{shantia2015indoor} &CNN &Real noisy image denoising &CNN with semi-supervised learning for 3D map\\
\hline
Choi et al. (2019) \cite{choi2019semi} &CNN &Real noisy image denoising &CNN with semi-supervised learning for medical noisy image denoising\\
\hline
\end{tabular}}
\label{tab:booktabs}
\end{table}\
\subsection{Deep learning techniques for blind denoising}
In the real world, images are easily corrupted and noise is complex. Therefore, blind denoising techniques are important \cite{loo2017survey}. An FFDNet \cite{zhang2018ffdnet} used noise level and noise as the input of CNN to train a denoiser for unknown noisy images. Subsequently, several methods were proposed to solve the problem of blind denoising. An image device mechanism proposed by Kenzo et al. \cite{isogawa2017deep} utilized soft shrinkage to adjust the noise level for blind denoising. For unpaired noisy images, using CNNs to estimate noise proved effective \cite{soltanayev2018training}. Yang et al. \cite{yang2017estimation} used known noise levels to train a denoiser, then utilized this denoiser to estimate the level of noise. To resolve the problem of random noise attenuation, a CNN with RL was used to filter complex noise \cite{zhang2018random,si2018random}. Changing the network architecture can improve the denoising performance for blind denoising. Majumdar et al. \cite{majumdar2018blind} proposed the use of an auto-encoder to tackle unknown noise. For mixed noise, cascaded CNNs were effective in removing the additive white Gaussian noise (AWGN) and impulse noise \cite{abiko2019blind}. Table 6 displays more information about these denoising methods.
\begin{table}[H]
\caption{Deep learning techniques for blind denoising.}
\label{tab:1}
\centering
\scalebox{0.50}[0.55]{
\begin{tabular}{|c|c|c|c|}
\hline
References &Methods &Applications &Key words (remarks)\\	
\hline
Zhang et al. (2018) \cite{zhang2018ffdnet}	&CNN	&Blind denoising	&CNN with varying noise level for blind denoising\\	
\hline
Kenzo et al. (2018) \cite{isogawa2017deep}	&CNN	&Blind denoising	&CNN with soft shrinkage for blind denoising\\
\hline
Soltanayev et al. (2018) \cite{soltanayev2018training}	&CNN	&Blind denoising	&CNN for unpaired noisy images\\
\hline
Yang et al. (2017) \cite{yang2017estimation}	&CNN	&Blind denoising	&CNNs with RL for blind denoising\\
\hline
Zhang et al. (2018) \cite{zhang2018random}	&CNN	&Blind denoising, random noise	&CNN with RL for blind denoising\\
\hline
Si et al. (2018) \cite{si2018random}	&CNN	&Blind denoising, random noise	&CNN for image denoising\\
\hline
Majumdar et al. (2018) \cite{jiang2018deep} 	&NN	&Blind denoising	&Auto-encoder for blind denoising\\
\hline
Abiko et al. (2019) \cite{godard2018deep}	&CNN	&Blind denoising, complex noisy image denoising	 &cascaded CNNs for blind denoising\\
\hline
Cha et al. (2019) \cite{yang2017estimation}	&CNN	&Blind denoising   &GAN for blind image denoising\\
\hline
Tian et al. (2020) \cite{tian2020attention} &CNN &Gaussian image denoising, blind denoising and real noisy image denoising &CNN with attention mechanism and sparse method for image denoising\\
\hline
\end{tabular}}
\label{tab:booktabs}
\end{table}\
\subsection{Deep learning techniques for hybrid noisy image denoising}
In the real world, captured images are affected by complex environments. Motivated by that, several researchers proposed hybrid-noisy-image denoising techniques. Li et al. \cite{li2018learning} proposed the combination of CNN and warped guidance to resolve the questions of noise, blur and JPEG compression. Zhang et al. \cite{zhang2018learning} used a model to deal with multiple degradations, such as noise, blur kernel and low-resolution image. To enhance the raw sensor data, Kokkinos et al. \cite{kokkinos2019iterative} presented a residual CNN with an iterative algorithm for image demosaicing and denoising. To handle arbitrary blur kernels, Zhang et al. \cite{zhang2019deep} proposed to use cascaded deblurring and single-image super-resolution (SISR) networks to recover plug-and-play super-resolution images. These hybrid noisy image denoising methods are presented in Table 7.
\begin{table}[H]
\caption{Deep learning techniques for hybrid noisy image denoising.}
\label{tab:1}
\centering
\scalebox{0.60}[0.60]{
\begin{tabular}{|c|c|c|c|}
\hline
References &Methods &Applications &Key words (remarks)\\	
\hline
Li  et al. (2018) \cite{li2018learning}	&CNN	&Noise, blur kernel, JPEG compression	&The combination of CNN and warped guidance for multiple degradations\\
\hline
Zhang et al. (2018) \cite{zhang2018learning}	&CNN	&Noise, blur kernel, low-resolution image	&CNN for multiple degradations\\
\hline
Kokkinos et al. (2019) \cite{kokkinos2019iterative} &CNN	&Image demosaicking and denoising	&Residual CNN with iterative algorithm for image demosaicking and denoising\\
\hline
\end{tabular}}
\label{tab:booktabs}
\end{table}\
It is noted that an image carries finite information, which is not beneficial in real-world applications. To address this problem, burst techniques were developed \cite{xia2019basis}. However, the burst image suffered from the effects of noise and camera shake, which increased the difficulty of implementing the actual task. Recently, there has been much interest in deep learning technologies for burst image denoising, where the noise is removed frame by frame \cite{aittala2018burst}. Recurrent fully convolutional deep neural networks can filter the noise for all frames in a sequence of arbitrary length \cite{godard2018deep}. The combination of CNN and the kernel method can boost the denoising performance for burst noisy images \cite{marinvc2019multi,mildenhall2018burst}. In terms of complex background noisy images, an attention mechanism combined the kernel and CNN to enhance the effect of key features for burst image denoising, which can accelerate the training speed \cite{zhang2020attention}. For low-light conditions, using a CNN to map a given burst noisy image to sRGB outputs can obtain a multi-frame denoising image sequence \cite{zhao2019end}. To reduce network complexity, a CNN with residual learning directly trained a denoising model rather than an explicit aligning procedure \cite{tan2019deep}. These burst denoising methods are listed in Table 8.
\begin{table}[H]
\caption{Deep learning techniques for burst denoising.}
\label{tab:1}
\centering
\scalebox{0.60}[0.60]{
\begin{tabular}{|c|c|c|c|}
\hline
References &Methods &Applications &Key words (remarks)\\	
\hline
Xia  et al. (2019) \cite{xia2019basis}	&CNN	&Burst denoising	&CNN for burst denoising\\
\hline
Aittala et al. (2018) \cite{aittala2018burst}	&CNN	&Burst denoising	&CNN for burst denoising\\
\hline
Godard et al. (2018) \cite{godard2018deep} &CNN	&Burst denoising	&CNN for burst denoising\\
\hline
Marin et al. (2019) \cite{marinvc2019multi} &CNN	&Burst denoising	&CNN with kernel idea for burst denoising\\
\hline
Mildenhall et al. (2018) \cite{mildenhall2018burst} &CNN	&Burst denoising	&CNN with kernel idea for burst denoising\\
\hline
Zhang et al. (2020) \cite{zhang2020attention} &CNN	&Burst denoising	&CNN with kernel idea and attention idea for burst denoising\\
\hline
Zhao et al. (2019) \cite{zhao2019end} &CNN	&Burst denoising	&CNN for burst denoising\\
\hline
Tan et al. (2019) \cite{tan2019deep} &CNN	&Burst denoising	&CNN without explicit aligning procedure for burst denoising\\
\hline
\end{tabular}}
\label{tab:booktabs}
\end{table}\

Similar to burst images, video detection is decomposed into each frame. Therefore, deep learning techniques for additive white noisy-image denoising, real noisy image denoising, blind denoising , hybrid noisy image denoising are also suitable to video denoising \cite{sadda2018real,wang2020first}. A recurrent neural network \cite{chen2016deep} utilized an end-to-end CNN to remove the noise from corrupted video. To improve video denoising, reducing the video redundancy is an effective method. A non-local patch idea fused CNN can efficiently suppress the noise for video and image denoising \cite{davy2018non}. A CNN combined temporal information to make a tradeoff between performance and training efficiency in video denoising \cite{tassano2019dvdnet}. For blind video denoising, a two-stage CNN proved to be a good choice \cite{ehret2019model}. The first phase trained a video denoising model by fine-tuning a pre-trained AWGN denoising network \cite{ehret2019model}. The second phase obtained latent clean video by the obtained video denoising model. These video denoising methods are described in Table 9.
\begin{table}[H]
\caption{Deep learning techniques for video denoising.}
\label{tab:1}
\centering
\scalebox{0.60}[0.60]{
\begin{tabular}{|c|c|c|c|}
\hline
References &Methods &Applications &Key words (remarks)\\	
\hline
Sadda  et al. (2018) \cite{sadda2018real}	&CNN	&Medical noisy video 	&CNN for video denoising\\
\hline
Wang et al. (2020) \cite{wang2020first}	&CNN	&Additive white Gaussian and salt-and-pepper noisy video 	&CNN for video denoising\\
\hline
Chen et al. (2016) \cite{chen2016deep} &CNN	&Additive white Poisson-Gaussian noisy video	&CNN for video denoising\\
\hline
Davy et al. (2018) \cite{davy2018non} &CNN	&Additive white Gaussian noisy video	&CNN with non-local idea for video denoising\\
\hline
Tassano et al. (2019) \cite{tassano2019dvdnet} &CNN	&Additive white Gaussian noisy video	&CNN with temporal information for video denoising\\
\hline
Ehret et al. (2019) \cite{ehret2019model} &CNN	&blind video denoising	&CNN with pre-trained technology for blind video denoising\\
\hline
\end{tabular}}
\label{tab:booktabs}
\end{table}\
\section{Experimental results}
\subsection{Datasets}
\subsubsection{Training datasets}
The training datasets are divided into two categories: gray-noisy and color-noisy images. Gary-noisy image datasets can be used to train Gaussian denoisers and blind denoisers. They included the BSD400 dataset \cite{bigdeli2017deep} and Waterloo Exploration Database \cite{ma2016waterloo}. The BSD400 dataset was composed of 400 images in .png format, and was cropped into a size of $180 \times 180$ for training a denoising model. The Waterloo Exploration Database consisted of 4,744 nature images with a .png format. Color-noisy images included the BSD432 \cite{zhang2017beyond}, Waterloo Exploration Database and polyU-Real-World-Noisy-Images datasets \cite{xu2018real}. Specifically, the polyU-Real-World-Noisy-Images consisted of 100 real noisy images with sizes of $2,784 \times 1,856$ obtained by five cameras: a Nikon D800, Canon 5D Mark II, Sony A7 II, Canon 80D and Canon 600D.
\subsubsection{Test datasets}
The test datasets included gray-noisy and color-noisy image datasets. The gray-noisy image dataset was composed of Set12 and BSD68 \cite{zhang2017beyond}. The Set12 contained 12 scenes. The BSD68 contained 68 nature images. They were used to test the Gaussian denoiser and a denoiser of blind noise. The color-noisy image dataset included CBSD68, Kodak24 \cite{franzen1999kodak}, McMaster \cite{zhang2011color}, cc \cite{nam2016holistic}, DND \cite{plotz2017benchmarking}, NC12 \cite{lebrun2015noise}, SIDD \cite{abdelhamed2018high} and Nam \cite{nam2016holistic}. The Kodak24 and McMaster contained 24 and 18 color noisy images, respectively. The cc contained 15 real noisy images of different ISO, i.e., 1,600, 3,200 and 6,400. The DND contained 50 real noisy images and the clean images were captured by low-ISO images. The NC12 contained 12 noisy images and did not have ground-truth clean images. The SIDD contained real noisy images from smart phones, and consisted of 320 image pairs of noisy and ground-truth images. The Nam included 11 scenes, which were saved in JPGE format.
\subsection{Experimental results}
To verify the denoising performance of some methods mentaioned in Section 3, we conducted some experiments on the Set12, BSD68, CBSD68, Kodak24, McMaster, DND, SIDD, Nam, cc and NC12 datasets in terms of quantitative and qualitative evaluations. The quantitative evaluation mainly used peak-signal-to-noise-ratio (PSNR) \cite{hore2010image} values of different denoisers to test the denoising effects. Additionally, we used the runtime of denoising of an image to support the PSNR for quantitative evaluation. The qualitative evaluation used visual figures to show the recovered clean images.
\begin{table}[t!]
\caption{PSNR (dB) of different methods on the BSD68 for different noise levels (i.e., 15, 25 and 50).}
\label{tab:1}
\centering
\scalebox{0.65}[0.65]{
\begin{tabular}{|c|c|c|c|c|}
\hline
Methods &15 &25 &50\\	
\hline
BM3D \cite{dabov2007image}	&31.07	&28.57	&25.62\\
\hline
WNNM \cite{gu2014weighted}	&31.37	&28.83	&25.87\\
\hline
EPLL \cite{zoran2011learning}	&31.21	&28.68	&25.67\\
\hline
MLP \cite{burger2012image}	&-	&28.96	&26.03\\
\hline
CSF \cite{schmidt2014shrinkage}	&31.24	&28.74	&-\\
\hline
TNRD \cite{chen2016trainable}	&31.42	&28.92	&25.97\\
\hline
ECNDNet	\cite{tian2019enhanced} &31.71	&29.22	&26.23\\
\hline
RED \cite{mao2016image}	&-	&-	&26.35\\
\hline
DnCNN \cite{zhang2017beyond}	&31.72	&29.23	&26.23\\
\hline
DDRN \cite{wang2017dilated}	&31.68	&29.18	&26.21\\
\hline
PHGMS \cite{bae2017beyond}	&31.86	&-	&26.36\\
\hline
MemNet \cite{tai2017memnet}	&-	&-	&26.35\\
\hline
EEDN \cite{chen2018imaged}	&31.58	&28.97	&26.03\\
\hline
NBCNN \cite{uchida2018non}	&31.57	&29.11	&26.16\\
\hline
NNC \cite{zarshenas2018deep}	&31.49	&28.88	&25.25\\
\hline
ELDRN \cite{panda2018exponential}	&32.11	&29.68	&26.76\\
\hline
PSN-K \cite{aljadaany2019proximal}	&31.70	&29.27	&26.32\\
\hline
PSN-U \cite{aljadaany2019proximal}	&31.60	&29.17	&26.30\\
\hline
DDFN \cite{couturier2018image}	&31.66	&29.16	&26.19\\
\hline
CIMM \cite{anwar2017chaining}	&31.81	&29.34	&26.40\\
\hline
DWDN \cite{li2018cnn}	&31.78	&29.36	&-\\
\hline
MWCNN \cite{liu2018multi}	&31.86	&29.41	&26.53\\
\hline
BM3D-Net \cite{yang2017bm3d} &31.42 &28.83	&25.73\\
\hline
MPFE-CNN \cite{kadimesetty2018convolutional} &31.79 &29.31 &26.34\\
\hline
IRCNN \cite{zhang2017learning}	&31.63	&29.15	&26.19\\
\hline
FFDNet \cite{zhang2018ffdnet}	&31.62	&29.19	&26.30\\
\hline
BRDNet \cite{tian2020image}	&31.79 &29.29 &26.36\\
\hline
ETN \cite{wang2017elu}	&31.82	&29.34	&26.32\\
\hline
ADNet \cite{tian2020attention} &31.74 &29.25 &26.29\\
\hline
NN3D \cite{cruz2018nonlocality}	&-	&-	&26.42\\
\hline
FOCNet \cite{jia2019focnet}	&31.83	&29.38	&26.50\\
\hline
DudeNet \cite{tian2020designing} &31.78 &29.29 &26.31\\
\hline
\end{tabular}}
\label{tab:booktabs}
\end{table}\
\begin{table}[t!]
\caption{FSIM of different methods on the BSD68 for different noise levels (i.e., 15, 25 and 50).}
\label{tab:1}
\centering
\scalebox{0.65}[0.65]{
\begin{tabular}{|c|c|c|c|c|}
\hline
Methods &15 &25 &50\\	
\hline
BM3D \cite{dabov2007image}	&0.9894	&0.9811	&0.9629\\
\hline
MLP \cite{burger2012image}	&0.9671	&0.9821	&0.9344\\
\hline
TNRD \cite{chen2016trainable}	&0.9697	&0.9820	&0.9291\\
\hline
ECNDNet	\cite{tian2019enhanced} &0.9911	&0.9837	&0.9686\\
\hline
IRCNN \cite{zhang2017learning}	&0.9905	&0.9835	&0.9700\\
\hline
BRDNet \cite{tian2020image}	&0.9913 &0.9841 &0.9687\\
\hline
ADNet \cite{tian2020attention} &0.9912 &0.9837 &0.9673\\
\hline
\end{tabular}}
\label{tab:booktabs}
\end{table}\
\subsubsection{Deep learning techniques for additive white noisy-image denoising}
Comparisons of denoising methods should take into consideration additive white noise, including Gaussian, Poisson, low-light noise, and salt and pepper noise, all of which have significantly different noise levels. Furthermore, many of the methods use different tools, which can have a significant influence on denoising results. For these reasons, we chose typical Gaussian noise to test the denoising performance of the various methods. In addition, most of the denoising methods use PSNR as a quantitative index. Therefore, we used the BSD68, Set12, CBSD68, Kodak24 and McMaster datasets to test the denoising performance of deep learning techniques for additive white noisy-image denoising. Table 10 shows the PSNR values of different networks with different noise levels for gray additive white noisy image denoising. To understand the denoising performance of different methods, we used a feature similarity index (FSIM) \cite{zhang2011fsim} as a visual quality metric to conduct experiments on BSD68 for different noise levels (i.e., 15, 25 and 50), as shown in Table 11.  To test the ability of dealing with single gray additive white noisy images from different networks, Set12 was used to conduct experiments, as shown in Table 12. Table 13 displays the denoising performance of different methods for color additive white noisy image denoising. Table 14 presents the efficiency of different methods for image denoising. For qualitative analysis, we magnified one area of the latent clean image from different methods. As shown in Figs. 7-10, the observed area is clearer, and the corresponding method has better denoising performance.
\begin{table}[t!]
\caption{PSNR (dB) of different methods on the Set12 for different noise levels (i.e., 15, 25 and 50).}
\label{tab:1}
\centering
\scalebox{0.60}[0.70]{
\begin{tabular}{|c|c|c|c|c|c|c|c|c|c|c|c|c|c|}
\hline
Images	&C.man	&House	&Peppers	&Starfish	&Monarch	&Airplane	&Parrot	&Lena	&Barbara	&Boat	&Man	&Couple	&Average\\	
\hline
Noise Level  & \multicolumn{13}{c|}{$\sigma$ = 15} \\
\hline
BM3D \cite{dabov2007image}	&31.91	&34.93	&32.69	&31.14	&31.85	&31.07	&31.37	&34.26	&33.10	&32.13	&31.92	&32.10	&32.37\\
\hline
WNNM \cite{gu2014weighted} &32.17	&35.13	&32.99	&31.82	&32.71	&31.39	&31.62	&34.27	&33.60	&32.27	&32.11	&32.17	&32.70\\
\hline
EPLL \cite{zoran2011learning}	&31.85	&34.17	&32.64	&31.13	&32.10	&31.19	&31.42	&33.92	&31.38	&31.93	&32.00	&31.93	&32.14\\
\hline
CSF \cite{schmidt2014shrinkage}	&31.95	&34.39	&32.85	&31.55	&32.33	&31.33	&31.37	&34.06	&31.92	&32.01	&32.08	&31.98	&32.32\\
\hline
TNRD \cite{chen2016trainable} &32.19	&34.53	&33.04	&31.75	&32.56	&31.46	&31.63	&34.24	&32.13	&32.14	&32.23	&32.11	&32.50\\
\hline
ECNDNet \cite{tian2019enhanced}	&32.56	&34.97	&33.25	&32.17	&33.11	&31.70	&31.82	&34.52	&32.41	&32.37	&32.39	&32.39	&32.81\\
\hline
DnCNN \cite{zhang2017beyond}	&32.61	&34.97	&33.30	&32.20	&33.09	&31.70	&31.83	&34.62	&32.64	&32.42	&32.46	&32.47	&32.86\\
\hline
PSN-K \cite{aljadaany2019proximal}	&32.58	&35.04	&33.23	&32.17	&33.11	&31.75	&31.89	&34.62	&32.64	&32.52	&32.39	&32.43	&32.86\\
\hline
PSN-U \cite{aljadaany2019proximal}	&32.04	&35.03	&33.21	&31.94	&32.93	&31.61	&31.62	&34.56	&32.49	&32.41	&32.37	&32.43	&32.72\\
\hline
CIMM \cite{anwar2017chaining}	&32.61	&35.21	&33.21	&32.35	&33.33	&31.77	&32.01	&34.69	&32.74	&32.44	&32.50	&32.52	&32.95\\
\hline
IRCNN \cite{zhang2017learning}	&32.55	&34.89	&33.31	&32.02	&32.82	&31.70	&31.84	&34.53	&32.43	&32.34	&32.40	&32.40	&32.77\\
\hline
FFDNet \cite{zhang2018ffdnet}	&32.43	&35.07	&33.25	&31.99	&32.66	&31.57	&31.81	&34.62	&32.54	&32.38	&32.41	&32.46	&32.77\\
\hline
BRDNet \cite{tian2020image} &32.80	&35.27	&33.47	&32.24	&33.35	&31.85	&32.00	&34.75	&32.93	&32.55	&32.50	&32.62	&33.03\\
\hline
ADNet \cite{tian2020attention} &32.81 &35.22 &33.49 &32.17 &33.17 &31.86 &31.96 &34.71 &32.80 &32.57 &32.47 &32.58 &32.98\\
\hline
DudeNet \cite{tian2020designing} &32.71 &35.13 &33.38 &32.29 &33.28 &31.78 &31.93 &34.66 &32.73 &32.46 &32.46 &32.49 &32.94\\
\hline
Noise Level  & \multicolumn{13}{c|}{$\sigma$ = 25} \\
\hline
BM3D \cite{dabov2007image}	&29.45	&32.85	&30.16	&28.56	&29.25	&28.42	&28.93	&32.07	&30.71	&29.90	&29.61	&29.71	&29.97\\
\hline
WNNM \cite{gu2014weighted} &29.64	&33.22	&30.42	&29.03	&29.84	&28.69	&29.15	&32.24	&31.24	&30.03	&29.76	&29.82	&30.26\\
\hline
EPLL \cite{zoran2011learning}	&29.26	&32.17	&30.17	&28.51	&29.39	&28.61	&28.95	&31.73	&28.61	&29.74	&29.66	&29.53	&29.69\\
\hline
MLP \cite{burger2012image}	&29.61	&32.56	&30.30	&28.82	&29.61	&28.82	&29.25	&32.25	&29.54	&29.97	&29.88	&29.73	&30.03\\
\hline
CSF \cite{schmidt2014shrinkage}	&29.48	&32.39	&30.32	&28.80	&29.62	&28.72	&28.90	&31.79	&29.03	&29.76	&29.71	&29.53	&29.84\\
\hline
TNRD \cite{chen2016trainable} &29.72	&32.53	&30.57	&29.02	&29.85	&28.88	&29.18	&32.00	&29.41	&29.91	&29.87	&29.71	&30.06\\
\hline
ECNDNet \cite{tian2019enhanced}	&30.11	&33.08	&30.85	&29.43	&30.30	&29.07	&29.38	&32.38	&29.84	&30.14	&30.03	&30.03	&30.39\\
\hline
DnCNN \cite{zhang2017beyond}	&30.18	&33.06	&30.87	&29.41	&30.28	&29.13	&29.43	&32.44	&30.00	&30.21	&30.10	&30.12	&30.43\\
\hline
PSN-K \cite{aljadaany2019proximal}	&30.28	&33.26	&31.01	&29.57	&30.30	&29.28	&29.38	&32.57	&30.17	&30.31	&30.10	&30.18	&30.53\\
\hline
PSN-U \cite{aljadaany2019proximal}	&29.79	&33.23	&30.90	&29.30	&30.17	&29.06	&29.25	&32.45	&29.94	&30.25	&30.05	&30.12	&30.38\\
\hline
CIMM \cite{anwar2017chaining}	&30.26	&33.44	&30.87	&29.77	&30.62	&29.23	&29.61	&32.66	&30.29	&30.30	&30.18	&30.24	&30.62\\
\hline
IRCNN \cite{zhang2017learning}	&30.08	&33.06	&30.88	&29.27	&30.09	&29.12	&29.47	&32.43	&29.92	&30.17	&30.04	&30.08	&30.38\\
\hline
FFDNet \cite{zhang2018ffdnet}	&30.10	&33.28	&30.93	&29.32	&30.08	&29.04	&29.44	&32.57	&30.01	&30.25	&30.11	&30.20	&30.44\\
\hline
BRDNet \cite{tian2020image} &31.39	&33.41	&31.04	&29.46	&30.50	&29.20	&29.55	&32.65	&30.34	&30.33	&30.14	&30.28	&30.61\\
\hline
ADNet \cite{tian2020attention} &30.34 &33.41 &31.14 &29.41 &30.39 &29.17 &29.49 &32.61 &30.25 &30.37 &30.08 &30.24 &30.58\\
\hline
DudeNet \cite{tian2020designing}  &30.23 &33.24 &30.98 &29.53 &30.44 &29.14 &29.48 &32.52 &30.15 &30.24 &30.08 &30.15 &30.52\\
\hline
Noise Level  & \multicolumn{13}{c|}{$\sigma$ = 50} \\
\hline
BM3D \cite{dabov2007image}	&26.13	&29.69	&26.68	&25.04	&25.82	&25.10	&25.90	&29.05	&27.22	&26.78	&26.81	&26.46	&26.72\\
\hline
WNNM \cite{gu2014weighted} &26.45	&30.33	&26.95	&25.44	&26.32	&25.42	&26.14	&29.25	&27.79	&26.97	&26.94	&26.64	&27.05\\
\hline
EPLL \cite{zoran2011learning}	&26.10	&29.12	&26.80	&25.12	&25.94	&25.31	&25.95	&28.68	&24.83	&26.74	&26.79	&26.30	&26.47\\
\hline
MLP \cite{burger2012image}	&26.37	&29.64	&26.68	&25.43	&26.26	&25.56	&26.12	&29.32	&25.24	&27.03	&27.06	&26.67	&26.78\\
\hline
TNRD \cite{chen2016trainable}	&26.62	&29.48	&27.10	&25.42	&26.31	&25.59	&26.16	&28.93	&25.70	&26.94	&26.98	&26.50	&26.81\\
\hline
ECNDNet \cite{tian2019enhanced}	&27.07	&30.12	&27.30	&25.72	&26.82	&25.79	&26.32	&29.29	&26.26	&27.16	&27.11	&26.84	&27.15\\
\hline
DnCNN \cite{zhang2017beyond}	&27.03	&30.00	&27.32	&25.70	&26.78	&25.87	&26.48	&29.39	&26.22	&27.20	&27.24	&26.90	&27.18\\
\hline
PSN-K \cite{aljadaany2019proximal}	&27.10	&30.34	&27.40	&25.84	&26.92	&25.90	&26.56	&29.54	&26.45	&27.20	&27.21	&27.09	&27.30\\
\hline
PSN-U \cite{aljadaany2019proximal}	&27.21	&30.21	&27.53	&25.63	&26.93	&25.89	&26.62	&29.54	&26.56	&27.27	&27.23	&27.04	&27.31\\
\hline
CIMM \cite{anwar2017chaining}	&27.25	&30.70	&27.54	&26.05	&27.21	&26.06	&26.53	&29.65	&26.62	&27.36	&27.26	&27.24	&27.46\\
\hline
IRCNN \cite{zhang2017learning}	&26.88	&29.96	&27.33	&25.57	&26.61	&25.89	&26.55	&29.40	&26.24	&27.17	&27.17	&26.88	&27.14\\
\hline
FFDNet \cite{zhang2018ffdnet}	&27.05	&30.37	&27.54	&25.75	&26.81	&25.89	&26.57	&29.66	&26.45	&27.33	&27.29	&27.08	&27.32\\
\hline
BRDNet \cite{tian2020image} &27.44	&30.53	&27.67	&25.77	&26.97	&25.93	&26.66	&29.73	&26.85	&27.38	&27.27	&27.17	&27.45\\
\hline
ADNet \cite{tian2020attention} &27.31 &30.59 &27.69 &25.70 &26.90 &25.88 &26.56 &29.59 &26.64 &27.35 &27.17 &27.07 &27.37\\
\hline
DudeNet \cite{tian2020designing} &27.22 &30.27 &27.51 &25.88 &26.93 &25.88 &26.50 &29.45 &26.49 &27.26 &27.19 &26.97 &27.30\\
\hline
\end{tabular}}
\label{tab:booktabs}
\end{table}\
\begin{table*}[t!]
\caption{ PSNR (dB) of different methods on the CBSD68, Kodak24 and McMaster for different noise levels (i.e., 15, 25, 35, 50 and 75).}
\label{tab:1}
\centering
\scalebox{0.65}[0.70]{
\begin{tabular}{|c|c|c|c|c|c|c|}
\hline
Datasets & Methods & $\sigma$ = 15 & $\sigma$ = 25 & $\sigma$ = 35 & $\sigma$ = 50 & $\sigma$ = 75 \\
\hline
\multirow{5}{*}{CBSD68}
&CBM3D \cite{dabov2007image}	&33.52	&30.71	&28.89	&27.38	&25.74\\
\cline{2-7} &DnCNN \cite{zhang2017beyond}	&33.98	&31.31	&29.65	&28.01	    &- \\
\cline{2-7} &DDRN \cite{wang2017dilated}	&33.93	&31.24	&-	&27.86	    &-\\
\cline{2-7} &EEDN \cite{chen2018imaged}	&33.65	&31.03	&-	&27.85	    &-\\
\cline{2-7}  &DDFN \cite{couturier2018image}	&34.17	&31.52	&29.88	&28.26	    &-\\
\cline{2-7}  &CIMM \cite{anwar2017chaining}	&31.81	&29.34	&-	&26.40	    &- \\
\cline{2-7} &BM3D-Net \cite{yang2017bm3d}	&33.79	&30.79	&-	&27.48	   &- \\
\cline{2-7} &IRCNN \cite{zhang2017learning} 	&33.86	&31.16	&29.50	&27.86	   &- \\
\cline{2-7} &FFDNet \cite{zhang2018ffdnet}	&33.80	&31.18	&29.57	&27.96	  &26.24 \\
\cline{2-7} &BRDNet \cite{tian2020image}	&34.10	&31.43	&29.77	&28.16	&26.43 \\
\cline{2-7} &GPADCNN \cite{cho2018gradient}	&33.83	&31.12	&29.46	&-	&- \\
\cline{2-7} &FFDNet \cite{tassano2019analysis}	&33.76	&31.18	&29.58	&-	&26.57 \\
\cline{2-7} &ETN \cite{wang2017elu}	&34.10	&31.41	&-	&28.01	&- \\
\cline{2-7} &ADNet \cite{tian2020attention} &33.99 &31.31 &29.66 &28.04 &26.33\\
\cline{2-7} &DudeNet \cite{tian2020designing} &34.01 &31.34 &29.71 &28.09 &26.40\\
\hline
\multirow{5}{*}{Kodak24}
&CBM3D \cite{dabov2007image}	&34.28	&31.68	&29.90	&28.46	&26.82\\
\cline{2-7} &DnCNN \cite{zhang2017beyond}	&34.73	&32.23	&30.64	&29.02	&-\\
\cline{2-7} &IRCNN \cite{zhang2017learning}	&34.56	&32.03	&30.43	&28.81	&-\\
\cline{2-7} &FFDNet \cite{zhang2018ffdnet}	&34.55	&32.11	&30.56	&28.99	&27.25\\
\cline{2-7} &BRDNet \cite{tian2020image}	&34.88	&32.41	&30.80	&29.22	&27.49\\
\cline{2-7} &FFDNet \cite{tassano2019analysis}	&34.53	&32.12	&30.59	&-	&27.61\\
\cline{2-7} &ADNet \cite{tian2020attention} &34.76 &32.26 &30.68 &29.10 &27.40\\
\cline{2-7} &DudeNet \cite{tian2020designing} &34.81 &32.26 &30.69 &29.10 &27.39\\
\hline
\multirow{5}{*}{McMaster}
&CBM3D \cite{dabov2007image}	&34.06	&31.66	&29.92	&28.51	&26.79 \\
\cline{2-7} &DnCNN \cite{zhang2017beyond}	&34.80	&32.47	&30.91	&29.21	&- \\
\cline{2-7} &IRCNN \cite{zhang2017learning}	&34.58	&32.18	&30.59 &28.91	&-\\
\cline{2-7} &FFDNet \cite{zhang2018ffdnet}	&34.47	&32.25	&30.76	&29.14	&27.29\\
\cline{2-7}  &BRDNet \cite{tian2020image}	&35.08	&32.75	&31.15	&29.52	&27.72\\
\cline{2-7} &ADNet \cite{tian2020attention} &34.93 &32.56 &31.00 &29.36 &27.53\\
\hline
\end{tabular}}
\end{table*}
\begin{table*}[t!]
\caption{Running time of 13 popular denoising methods for the noisy images of sizes 256 $\times$ 256, 512 $\times$ 512 and 1024 $\times$ 1024.}
\label{tab:1}
\centering
\scalebox{0.75}[0.75]{
\begin{tabular}{|c|c|c|c|c|}
\hline
Methods &Device &256 $\times$ 256 &512 $\times$ 512 &1024 $\times$ 1024\\
\hline
BM3D \cite{dabov2007image}	&CPU	&0.65	&2.85	&11.89 \\
\hline
WNNM \cite{gu2014weighted} 	&CPU	&203.1	&773.2	&2536.4\\
\hline
EPLL \cite{zoran2011learning}	&CPU	&25.4	&45.5	&422.1\\
\hline
MLP \cite{burger2012image}	&CPU	&1.42	&5.51	&19.4 \\
\hline
CSF \cite{schmidt2014shrinkage}	&CPU	&2.11	&5.67	&40.8\\
\hline
CSF \cite{schmidt2014shrinkage}	&GPU &-	&0.92	&1.72 \\
\hline
TNRD \cite{chen2016trainable}	&CPU	&0.45	&1.33	&4.61\\
\hline
TNRD \cite{chen2016trainable}	&GPU	&0.010	&0.032	&0.116\\
\hline
ECNDNet \cite{tian2019enhanced}	&GPU	&0.012	&0.079	&0.205\\
\hline
DnCNN \cite{zhang2017beyond}	&CPU	&0.74	&3.41	&12.1\\
\hline
DnCNN \cite{zhang2017beyond}	&GPU	&0.014	&0.051	&0.200\\
\hline
FFDNet \cite{zhang2018ffdnet}	&CPU	&0.90	&4.11	&14.1\\
\hline
FFDNet \cite{zhang2018ffdnet}	&GPU	&0.016	&0.060	&0.235\\
\hline
IRCNN \cite{zhang2017learning}	&CPU	&0.310	&1.24	&4.65\\
\hline
IRCNN \cite{zhang2017learning}	&GPU	&0.012	&0.038	&0.146\\
\hline
BRDNet \cite{tian2020image}	&GPU	&0.062	&0.207	&0.788\\
\hline
ADNet \cite{tian2020attention} &GPU &0.0467 &0.0798 &0.2077\\
\hline
DudeNet \cite{tian2020designing} &GPU &0.018 &0.422 &1.246\\
\hline
\end{tabular}}
\end{table*}
\begin{figure}[!htb]
\centering
\subfloat{\includegraphics[width=4.5in]{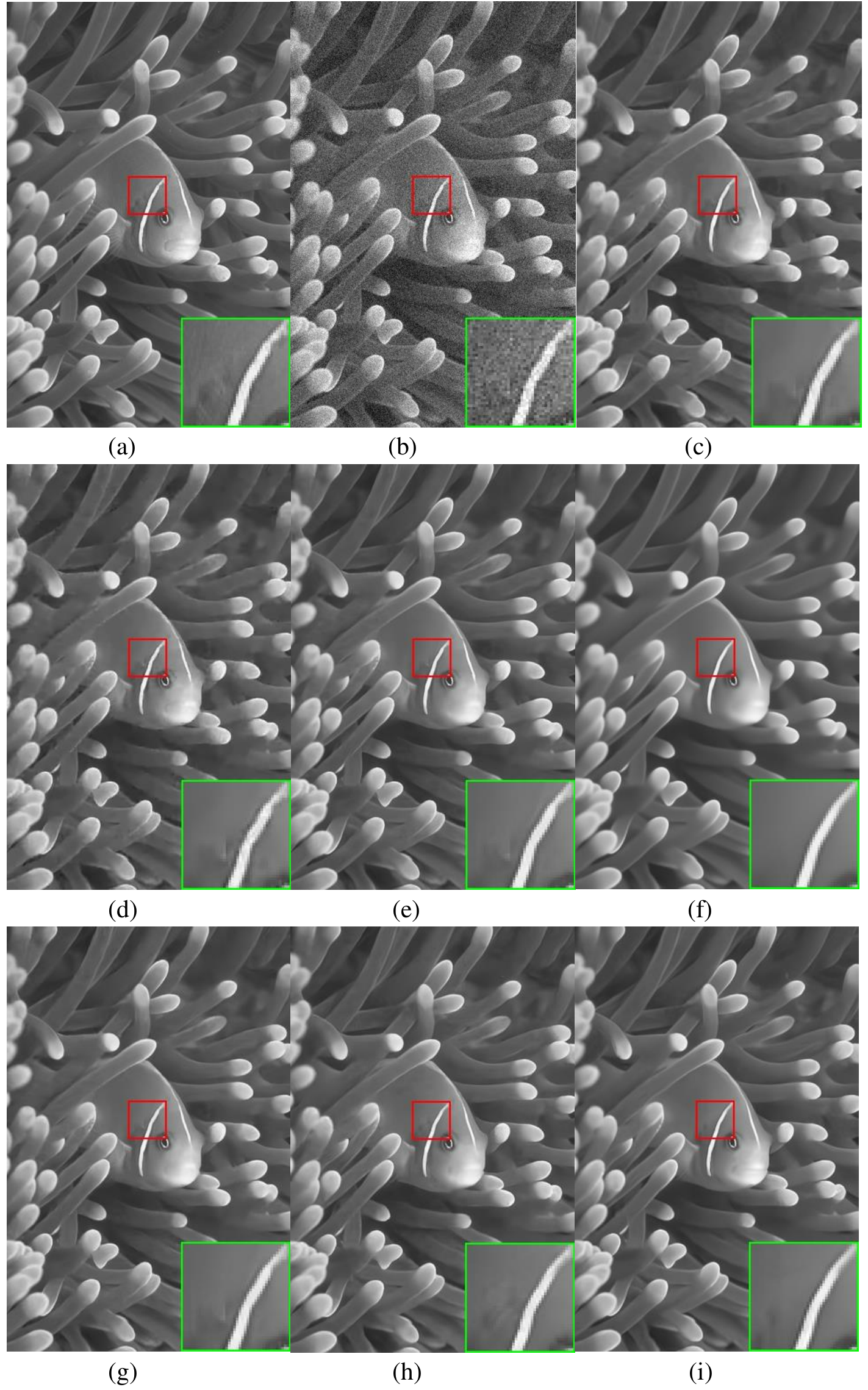}
\label{fig_second_case}}
\caption{Denoising results of different methods on one image from the BSD68 with $\sigma {\rm{ = }}15$: (a) original image, (b) noisy image/24.62dB, (c) BM3D/35.29dB, (d) EPLL/34.98dB, (e)
DnCNN/36.20dB, (f) FFDNet/36.75dB, (g) IRCNN/35.94dB, (h) ECNDNet/36.03dB, and (i) BRDNet/36.59dB.}
\label{fig:5}
\end{figure}
\begin{figure}[!htb]
\centering
\subfloat{\includegraphics[width=3.4in]{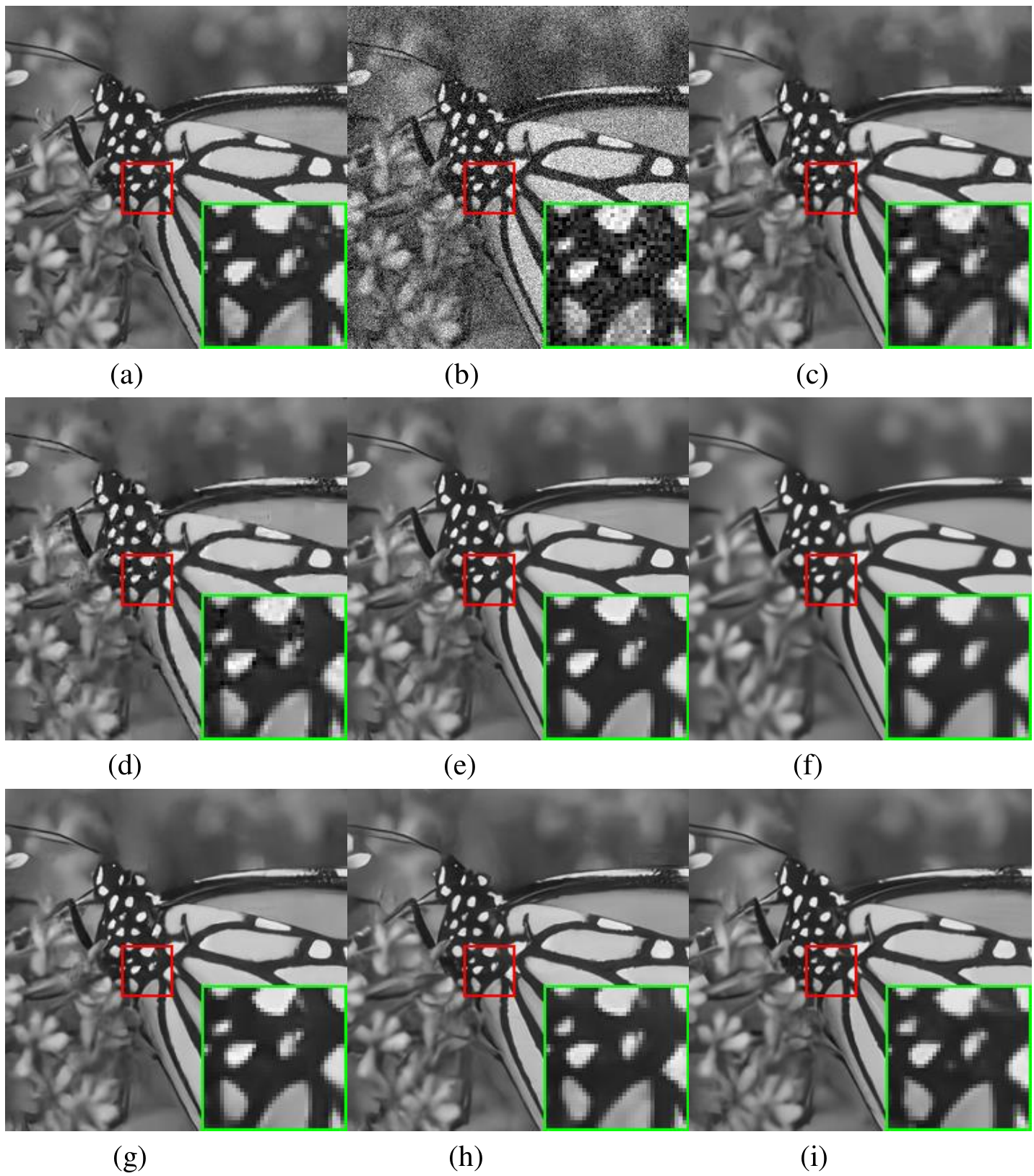}
\label{fig_second_case}}
\caption{Denoising results of different methods on one image from the Set12 with $\sigma {\rm{ = }}25$: (a) original image, (b) noisy image/20.22dB, (c) BM3D/29.26dB, (d) EPLL/29.44dB, (e)
DnCNN/30.28dB, (f) FFDNet/30.08dB, (g) IRCNN/30.09dB, (h) ECNDNet/30.30dB, and (i) BRDNet/30.50dB.}
\label{fig:5}
\end{figure}
\begin{figure}[!htb]
\centering
\subfloat{\includegraphics[width=4in]{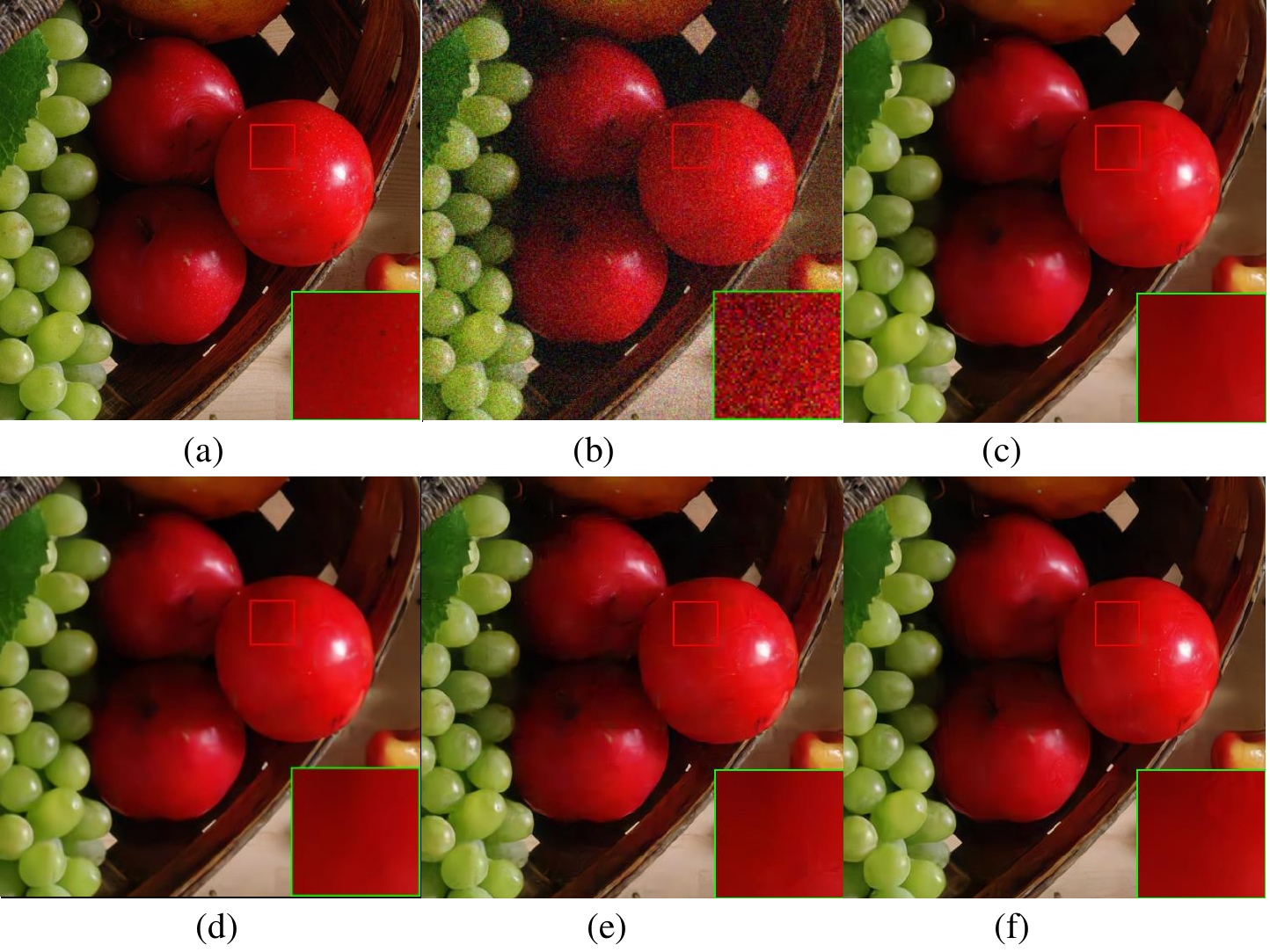}
\label{fig_second_case}}
\caption{Denoising results of different methods on one image from the McMaster with $\sigma {\rm{ = }}35$: (a) original image, (b) noisy image/18.46dB, (c) DnCNN/33.05B, (d) FFDNet/33.03dB, (e) IRCNN/32.74dB, and (f) BRDNet/33.26dB.}
\label{fig:5}
\end{figure}
\begin{figure}[!htb]
\centering
\subfloat{\includegraphics[width=4.5in]{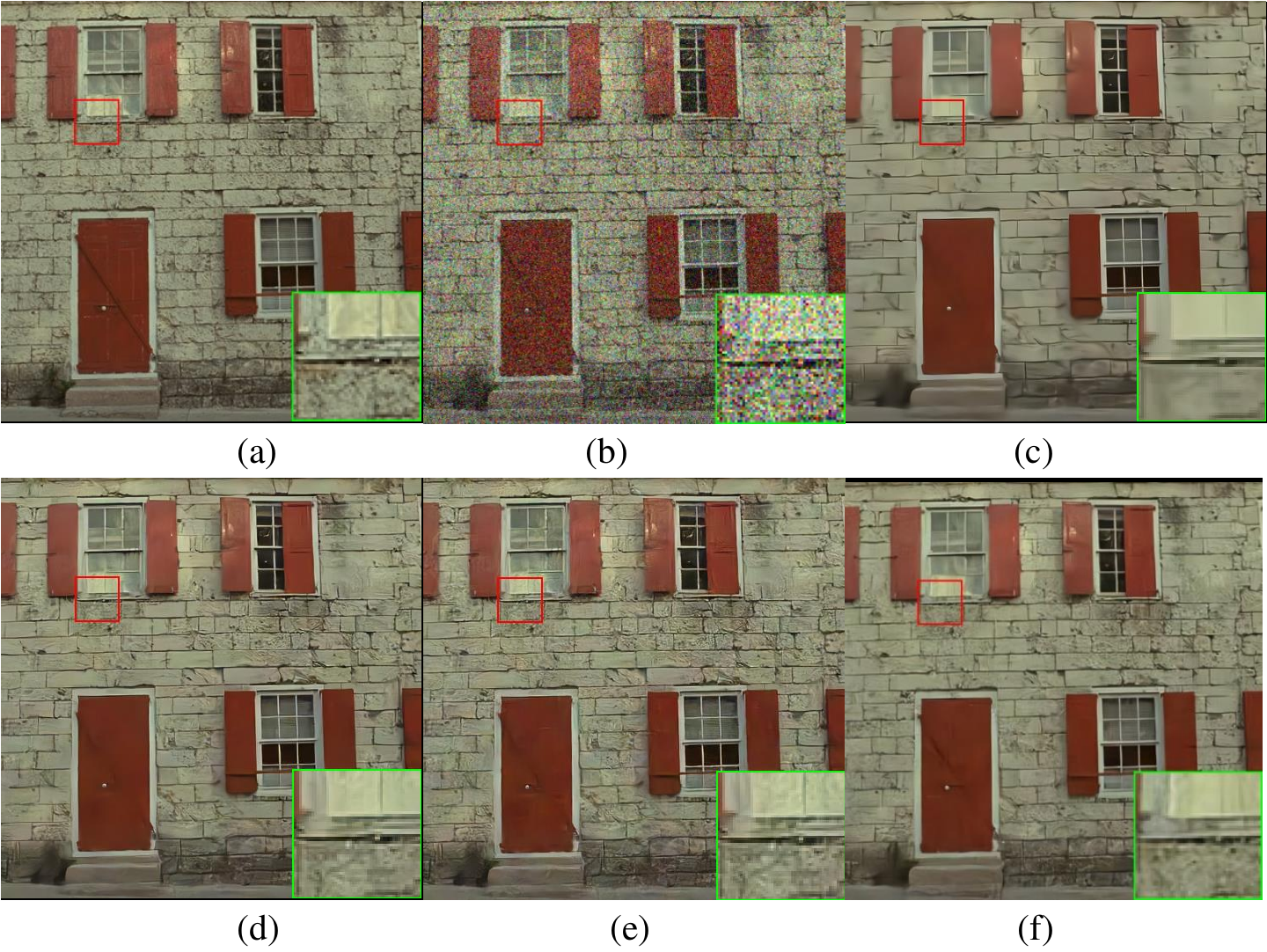}
\label{fig_second_case}}
\caption{ Denoising results of different methods on one image from the Kodak24 with $\sigma {\rm{ = }}50$: (a) original image, (b) noisy image/14.58dB, (c) DnCNN/25.80B, (d) FFDNet/26.13dB, (e) IRCNN/26.10dB, and (f) BRDNet/26.33dB.}
\label{fig:5}
\end{figure}
\subsubsection{Deep learning techniques for real-noisy image denoising}
For testing the denoising performance of deep learning techniques for real-noisy images, the public datasets, such as DND, SIDD, Nam and CC, were chosen to design the experiments. We chose not to use the NC12 dataset because the ground-truth clean images from NC12 were unavailable. Also, to help readers better understand these methods, we added several traditional denoising methods, such as BM3D, as comparative methods. From Tables 15 and 16, we can see that the DRDN obtained the best results on the DND and SSID in real-noisy image denoising, respectively. For compressed noisy images, the AGAN obtained excellent performance, as shown in Table 17. For real noisy images of different ISO values, the SDNet and BRDNet achieved the best and second-best denoising performance, respectively, as described in Table 18.
\begin{table*}[t!]
\caption{PSNR (dB) of different methods on the DND for real-noisy image denoising.}
\label{tab:1}
\centering
\scalebox{0.75}[0.75]{
\begin{tabular}{|c|c|}
\hline
Methods	&DND\\
\hline
EPLL \cite{zoran2011learning}	&33.51\\
\hline
TNRD \cite{chen2016trainable}	&33.65\\
\hline
NCSR \cite{dong2012nonlocally}	&34.05\\
\hline
MLP \cite{burger2012image}	&34.23\\
\hline
BM3D \cite{dabov2007image}	&34.51\\
\hline
FoE \cite{roth2005fields}	&34.62\\
\hline
WNNM \cite{gu2014weighted} 	&34.67\\
\hline
KSVD \cite{aharon2006k}	&36.49\\
\hline
CDnCNN-B \cite{zhang2017beyond}	&32.43\\
\hline
FFDNet \cite{zhang2018ffdnet}	&34.40\\
\hline
MCWNNM \cite{liu2018multi}	&37.38\\
\hline
TWSC \cite{xu2018trilateral}	&37.94\\
\hline
GCBD \cite{chen2018image}	&35.58\\
\hline
CIMM \cite{anwar2017chaining}	&36.04\\
\hline
CBDNet \cite{guo2019toward}	&37.72\\
\hline
VDN \cite{yue2019variational}	&39.38\\
\hline
DRDN \cite{song2019dynamic}	&39.40\\
\hline
AGAN \cite{lin2019real}		&38.13\\
\hline
\end{tabular}}
\end{table*}
\begin{table*}[t!]
\caption{PSNR (dB) of different methods on the SIDD for real-noisy image denoising.}
\label{tab:1}
\centering
\scalebox{0.75}[0.75]{
\begin{tabular}{|c|c|}
\hline
Methods	&SIDD\\
\hline
CBM3D \cite{dabov2007image}	&25.65\\
\hline
WNNM \cite{gu2014weighted} 	&25.78\\
\hline
MLP \cite{burger2012image}	&24.71\\
\hline
DnCNN-B \cite{zhang2017beyond}	&23.66\\
\hline
CBDNet \cite{guo2019toward}	&33.28\\
\hline
VDN \cite{yue2019variational}	&39.23\\
\hline
DRDN \cite{song2019dynamic}	&39.60\\
\hline
\end{tabular}}
\end{table*}
\begin{table*}[t!]
\caption{PSNR (dB) of different methods on the Nam for real-noisy image denoising.}
\label{tab:1}
\centering
\scalebox{0.75}[0.75]{
\begin{tabular}{|c|c|}
\hline
Methods	&Nam\\
\hline
NI \cite{absoft2017neat}	&31.52\\
\hline
TWSC \cite{xu2018trilateral}	&37.52\\
\hline
BM3D \cite{dabov2007image}	&39.84\\
\hline
NC \cite{lebrun2015noise}	&40.41\\
\hline
WNNM \cite{gu2014weighted} 	&41.04\\
\hline
CDnCNN-B \cite{zhang2017beyond}	&37.49\\
\hline
MCWNNM \cite{liu2018multi}	&37.91\\
\hline
CBDNet \cite{guo2019toward}	&41.02\\
\hline
CBDNet(JPEG) \cite{guo2019toward}	&41.31\\
\hline
DRDN \cite{song2019dynamic}	&38.45\\
\hline
AGAN \cite{lin2019real}	&41.38\\
\hline
\end{tabular}}
\end{table*}
\begin{table*}[t!]
\caption{PSNR (dB) of different methods on the cc for real-noisy image denoising.}
\label{tab:1}
\centering
\scalebox{0.50}[0.60]{
\begin{tabular}{|c|c|c|c|c|c|c|c|c|c|c|c|}
\hline
Camera Settings &CBM3D \cite{dabov2007image} &MLP \cite{burger2012image}	&TNRD \cite{chen2016trainable}	&DnCNN \cite{zhang2017beyond}	&NI \cite{absoft2017neat}	&NC \cite{lebrun2015noise}	&WNNM \cite{gu2014weighted} 	&BRDNet \cite{tian2020image}	&SDNet \cite{zhao2019simple} &ADNet \cite{tian2020attention} &DudeNet \cite{tian2020designing}\\
\hline
\multirow{3}{*}{Canon 5D ISO=3200}
&39.76	&39.00	&39.51	&37.26	&35.68	&38.76	&37.51	&37.63	&39.83 &35.96 &36.66\\
\cline{2-12} &36.40	&36.34	&36.47	&34.13	&34.03	&35.69	&33.86	&37.28	&37.25 &36.11 &36.70\\
\cline{2-12} &36.37	&36.33	&36.45	&34.09	&32.63	&35.54	&31.43	&37.75	&36.79 &34.49 &35.03\\
\hline
\multirow{3}{*}{Nikon D600 ISO=3200}
&34.18	&34.70	&34.79	&33.62	&31.78	&35.57	&33.46	&34.55	&35.50 &33.94 &33.72\\
\cline{2-12} &35.07	&36.20	&36.37	&34.48	&35.16	&36.70	&36.09	&35.99	&37.24 &34.33 &34.70\\
\cline{2-12} &37.13	&39.33	&39.49	&35.41	&39.98	&39.28	&39.86	&38.62	&41.18 &38.87 &37.98\\
\hline
\multirow{3}{*}{Nikon D800 ISO=1600}
&36.81	&37.95	&38.11	&35.79	&34.84	&38.01	&36.35	&39.22	&38.77 &37.61 &38.10\\	
\cline{2-12} &37.76	&40.23	&40.52	&36.08	&38.42	&39.05	&39.99	&39.67	&40.87 &38.24 &39.15\\	
\cline{2-12} &37.51	&37.94	&38.17	&35.48	&35.79	&38.20	&37.15	&39.04	&38.86 &36.89 &36.14\\	
\hline
\multirow{3}{*}{Nikon D800 ISO=3200}
&35.05	&37.55	&37.69	&34.08	&38.36	&38.07	&38.60	&38.28	&39.94 &37.20 &36.93\\
\cline{2-12} &34.07	&35.91	&35.90	&33.70	&35.53	&35.72	&36.04	&37.18	&36.78 &35.67 &35.80\\
\cline{2-12} &34.42	&38.15	&38.21	&33.31	&40.05	&36.76	&39.73	&38.85	&39.78 &38.09 &37.49\\
\hline
\multirow{3}{*}{Nikon D800 ISO=6400}
&31.13	&32.69	&32.81	&29.83	&34.08	&33.49	&33.29	&32.75	&33.34 &32.24 &31.94\\
\cline{2-12} &31.22	&32.33	&32.33	&30.55	&32.13	&32.79	&31.16	&33.24	&33.29 &32.59 &32.51\\
\cline{2-12} &30.97	&32.29	&32.29	&30.09	&31.52	&32.86	&31.98	&32.89	&33.22 &33.14 &32.91\\
\hline
Average &35.19	&36.46	&36.61	&33.86	&35.33	&36.43	&35.77	&36.73	&37.51 &35.69 &35.72\\
\hline
\end{tabular}}
\end{table*}
\subsubsection{Deep learning techniques for blind denoising}
It is known that noise is complex in the real world, and not subject to rules. This is why blind denoising techniques, especially deep learning techniques, have been developed. Comparing the denoising performance of different deep learning techniques is very useful. The state-of-the-art denoising methods such as DnCNN, FFDNet, ADNet, SCNN and G2G1 on the BSD68 and Set12 were chosen to design the experiments. FFDNet and ADNet are superior to other methods in blind denoising, as shown in Tables 19 and 20, respectively.
\begin{table*}[t!]
\caption{Different methods on the BSD68 for different noise levels (i.e., 15, 25 and 50).}
\label{tab:1}
\centering
\scalebox{0.75}[0.75]{
\begin{tabular}{|c|c|c|c|}
\hline
Methods	&15	&25	&50\\
\hline
DnCNN-B \cite{zhang2017beyond}	&31.61	&29.16	&26.23\\
\hline
FFDNet \cite{zhang2018ffdnet}	&31.62	&29.19	&26.30\\
\hline
SCNN \cite{isogawa2017deep}	&31.48	&29.03	&26.08\\
\hline
ADNet-B \cite{tian2020attention} &31.56 &29.14 &26.24\\
\hline
DnCNN-SURE-T \cite{soltanayev2018training}	&-	&29.00	&25.95\\
\hline
DnCNN-MSE-GT \cite{soltanayev2018training}	&-	&29.20	&26.22\\
\hline
G2G1(LM,BSD) \cite{cha2019gan2gan}	&31.55	&28.93	&25.73\\
\hline
\end{tabular}}
\end{table*}
\begin{table*}[t!]
\caption{Average PSNR (dB) results of different methods on Set12 with noise levels of 25 and 50.}
\label{tab:1}
\centering
\scalebox{0.63}[0.63]{
\begin{tabular}{|c|c|c|c|c|c|c|c|c|c|c|c|c|c|}
\hline
Images & C.man & House & Peppers & Starfish & Monarch & Airplane & Parrot & Lena & Barbara & Boat & Man & Couple & Average \\
\hline
Noise Level  & \multicolumn{13}{c|}{$\sigma$ = 25} \\
\hline
DnCNN-B \cite{zhang2017beyond}	&29.94	&33.05	&30.84	&29.34	&30.25	&29.09	&29.35	&32.42	&29.69	&30.20	&30.09	&30.10	&30.36\\
\hline
FFDNet \cite{zhang2018ffdnet}	&30.10	&33.28	&30.93	&29.32	&30.08	&29.04	&29.44	&32.57	&30.01	&30.25	&30.11	&30.20	&30.44\\
\hline
ADNet-B \cite{tian2020attention} &29.94 &33.38 &30.99 &29.22 &30.38 &29.16 &29.41 &32.59 &30.05 &30.28 &30.01 &30.15 &30.46\\
\hline
DudeNet-B \cite{tian2020designing} &30.01 &33.15 &30.87 &29.39 &30.31 &29.07 &29.40 &32.42 &29.76 &30.18 &30.03 &30.06 &30.39\\
\hline
DNCNN-SURE-T \cite{soltanayev2018training}	&29.86	&32.73	&30.57	&29.11	&30.13	&28.93	&29.26	&32.08	&29.44	&29.86	&29.91	&29.78	&30.14\\
\hline
DNCNN-MSE-GT \cite{soltanayev2018training}	&30.14	&33.16	&30.84	&29.40	&30.45	&29.11	&29.36	&32.44	&29.91	&30.11	&30.08	&30.06	&30.42\\
\hline
Noise Level  & \multicolumn{13}{c|}{$\sigma$ = 50} \\
\hline
DnCNN-B \cite{zhang2017beyond}	&27.03	&30.02	&27.39	&25.72	&26.83	&25.89	&26.48	&29.38	&26.38	&27.23	&27.23	&26.91	&27.21\\
\hline
FFDNet \cite{zhang2018ffdnet}	&27.05	&30.37	&27.54	&25.75	&26.81	&25.89	&26.57	&29.66	&26.45	&27.33	&27.29	&27.08	&27.32\\
\hline
ADNet-B \cite{tian2020attention} &27.22 &30.43 &27.70 &25.63 &26.92 &26.03 &26.56 &29.53 &26.51 &27.22 &27.19 &27.05 &27.33\\
\hline
DudeNet-B \cite{tian2020designing} &27.19 &30.11 &27.50 &25.69 &26.82 &25.85 &26.46 &29.35 &26.38 &27.20 &27.13 &26.90 &27.22\\
\hline
DNCNN-SURE-T \cite{soltanayev2018training}	&26.47	&29.20	&26.78	&25.39	&26.53	&25.65	&26.21	&28.81	&25.23	&26.79	&26.97	&26.48	&26.71\\
\hline
DNCNN-MSE-GT \cite{soltanayev2018training}	&27.03	&29.92	&27.27	&25.65	&26.95	&25.93	&26.43	&29.31	&26.17	&27.12	&27.22	&26.94	&27.16\\
\hline
\end{tabular}}
\end{table*}
\subsubsection{Deep learning techniques for hybrid-noisy-image denoising}
In the real world, corrupted images may include different kinds of noise \cite{he2019multi}, which makes it very difficult to recover a latent clean image. To resolve this problem, deep learning techniques based multi-degradation idea have been proposed, as discussed in Section 3.4. Here we introduce the denoising performance of the multi-degradation model, as shown in Table 21, where the WarpNet method is shown to be very competitive in comparison with other popular denoising methods, such as DnCNN and MemNet.

\begin{table}[t!]
\caption{Different methods on the VggFace2and WebFace for image denoising.}
\label{tab:1}
\centering
\scalebox{0.7}[0.7]{
\begin{tabular}{|c|c|c|c|c|}
\hline
\multirow{2}{*}{Methods} &
\multicolumn{2}{c|}{VggFace2 \cite{cao2018vggface2}} &
\multicolumn{2}{c|}{WebFace \cite{yi2014learning}} \\
\cline{2-5}
 &4 $\times$   &8 $\times$  &4 $\times$   &8 $\times$\\
\hline
DnCNN \cite{zhang2017beyond} &26.73	&23.29	&28.35	&24.75\\
\hline
MemNet \cite{tai2017memnet} &26.85	&23.31	&28.57	&24.77\\
\hline
WarpNet \cite{li2018learning} &28.55 &24.10	&32.31	&27.21\\
\hline
\end{tabular}}
\label{tab:booktabs}
\end{table}
\section{Discussion}
Deep learning techniques are seeing increasing use in image denoising. This paper offers a survey of these techniques in order to help readers understand these methods. In this section, we present the potential areas of further research for image denoising and points out several as yet unsolved problems.

Image denoising based on deep learning techniques mainly are effective in increasing denoising performance and efficiency, and performing complex denoising tasks. Solutions for improving denoising performance include the following:

1) Enlarging the receptive field can capture more context information. Enlarging the receptive field can be accomplished by increasing the depth and width of the networks. However, this results in higher computational costs and more memory consumption. One technique for resolving this problem, is dilated convolution, which not only contributions to higher performance and efficiency, but is also very effective for mining more edge information.

2) The simultaneous use of extra information (also called prior knowledge) and a CNN is an effective approach to facilitate obtaining more accurate features. This is implemented by designing the loss function.

3) Combining local and global information can enhance the memory abilities of the shallow layers on deep layers to better filter the noise. Two methods for addressing this  problem are residual operation and recursive operation.

4) Single processing methods can be used to suppress the noise. The single processing technique fused into the deep CNN can achieve excellent performance. For example, the wavelet technique is gathered into the U-Net to deal with image restoration \cite{liu2018multi}.

5) Data augmentation, such as horizontal flip, vertical flip and color jittering, can help the denoising methods learn more types of noise, which can enhance the expressive ability of the denoising models. Additionally, using the GAN to construct virtual noisy images is also useful for image denoising.

6) Transfer learning, graph and neural architecture search methods can obtain good denoising results.

7) Improving the hardware or camera mechanism can reduce the effect of noise on the captured image.

Compressing deep neural networks has achieved great success in improving the efficiency of denoising. Reducing the depth or the width of deep neural networks can reduce the complexity of these networks in image denoising. Also, the use of small convolutional kernel and group convolution can reduce the number of parameters, thereby accelerating the speed of training. The fusion of dimension reduction methods, such as principal component analysis (PCA) and CNN, can also lead to improvements in denoising efficiency.

For resolving complex noisy images, step-by-step processing is a very popular method. For example, using a two-step mechanism is a way of dealing with a noisy image with low-resolution. The first step involves the recovery of a high-resolution image by a CNN. The second step uses a novel CNN to filter the noise of the high-resolution image.
In the example above, the two CNNs are implemented via a cascade operation. This two-step mechanism is ideal for unsupervised noise tasks, such as real noisy images and blind denoising. That is, the first step relies on a CNN with optimization algorithms, i.e., maximum a posteriori, to estimate the noise as ground truth (referred as a label). The second step utilized another CNN and obtained ground truth to train a denoising model for real-noisy image denoising or blind denoising. The self-supervised learning fused into the CNN is a good choice for real-noisy image denoising or blind denoising.

Although deep learning techniques have attained great success in these three scenarios, there are still challenges in the field of image denoising. These include:

1)	Deeper denoising networks require more memory resources.

2)	Training deeper denoising networks is not a stable solution for real noisy image, unpaired noisy image and multi-degradation tasks.

3)	Real noisy images are not easily captured, which results in inadequate training samples.

4)	Deep CNNs are difficult to solve unsupervised denoising tasks.

5)	More accurate metrics need to be found for image denoising. PSNR and SSIM are popular metrics for the task of image restoration. PSNR suffers from excessive smoothing, which is very difficult to recognize indistinguishable images. SSIM depends on brightness, contrast and structure, and therefore cannot accurately evaluate image perceptual quality.
\section{Conclusion}
In this paper, we compare, study and summarize the deep networks used for on image denoising. First, we show the basic frameworks of deep learning for image denoising. Then, we present the deep learning techniques for noisy tasks, including additive white noisy images, blind denoising, real noisy images and hybrid noisy images. Next, for each category of noisy tasks, we analyze the motivation and theory of denoising networks. Next, we compare the denoising results, efficiency and visual effects of different networks on benchmark datasets, and then perform a cross-comparison of the different types of image denoising methods with different types of noise. Finally, some potential areas for further research are suggested, and the challenges of deep learning in image denoising are discussed.

Over the past few years, Gaussian noisy image denoising techniques have achieved great success, particularly in scenarios where the Gaussian noise is regular. However, in the real world the noise is complex and irregular. Improving the hardware devices in order to better suppress the noise for capturing a high-quality image is very important. Moreover, the obtained image may be blurry, low-resolution and corrupted. Therefore, it is critical to determine how to effectively recover the latent clean image from the superposed noisy image. Furthermore, while the use of deep learning techniques to learn features requires the ground truth, the obtained real noisy images do not have the ground truth. These are urgent challenges that researches and scholars need to address.
\section*{Acknowledgments}
This paper is partially supported by the National Natural Science Foundation of China under Grant No. 61876051, in part by Shenzhen Municipal Science and Technology Innovation, Council under Grant No. JSGG20190220153602271 and in part by the Natural Science Foundation of Guang dong Province under Grant No. 2019A1515011811.
\section*{References}


\bibliographystyle{elsarticle-harv}
\bibliography{references}

\end{spacing}
\end{document}